\documentclass[11pt,a4paper]{article}
\usepackage{graphicx}
\usepackage{amssymb}
\usepackage{amsmath}
\usepackage{bm}
\usepackage{color}

\usepackage{subfigure}
\usepackage{caption}
\usepackage{feynmp}

\usepackage[compat=1.1.0]{tikz-feynhand}
\usepackage[compat=1.1.0]{tikz-feynman}
\usepackage{tikz}
\usepackage[sort&compress,numbers, merge]{natbib}

\usepackage{orcidlink}

\usepackage{dcolumn}

\usepackage{xcolor}
\usepackage{hyperref}
\usepackage[mathlines]{lineno}
\begin{document}

\begin{titlepage}
\begin{center}
\hfill UMN--TH--4122/22\\
\hfill FTPI-MINN-22/13

\vspace{2.0cm}
{\Large\bf Enhanced EDMs from Small Instantons}

\vspace{1.0cm}
{\bf Ravneet S. Bedi$^{a,}$\footnote{bedi0019@umn.edu}\orcidlink{0000-0002-7104-1753}, Tony Gherghetta$^{a,}$\footnote{tgher@umn.edu}\orcidlink{0000-0002-8489-1116}
and Maxim Pospelov$^{a,b,}$\footnote{pospelov@umn.edu}}

\vspace{0.5cm}
{\small\it
${}^a$ School of Physics and Astronomy, University of Minnesota, \\Minneapolis, Minnesota 55455, USA\\
${}^b$ William I.~Fine Theoretical Physics Institute, University of Minnesota,\\ Minneapolis, Minnesota 55455 USA}

\vspace{0.5cm}
\abstract
We show that models in which the strong $CP$ problem is solved by introducing an axion field with a mass 
enhanced by non-QCD UV dynamics at a scale  $\Lambda_{\rm SI}$
exhibit enhanced sensitivity to external sources of $CP$ violation. In the presence of higher-dimensional $CP$-odd sources at a scale $\Lambda_{\rm CP}$, the same mechanisms that enhance the axion mass also modify the axion potential, shifting the potential minimum by a factor $\propto\Lambda^2_{\rm SI}/\Lambda^2_{\rm CP}$.
This phenomenon of $CP$ violation enhancement, which puts stringent constraints on the scale of new physics, is explicitly demonstrated within a broad class of ``small instanton" models with
$CP$-odd sources arising from the dimension-six Weinberg gluonic and four-fermion operators. We find that for heavy axion masses $\gtrsim 100$\,MeV, 
arising from new dynamics at $\Lambda_{\rm SI}\lesssim 10^{10}$~GeV, 
$CP$ violation generated up to the Planck scale can be probed by future electric dipole moment experiments.

\flushbottom

\end{center}
    
\end{titlepage}

\setcounter{footnote}{0}

\section{Introduction}\label{Sec1}

The Standard Model (SM) of particle physics has two sources of $CP$ violation. The well-established and measured source of $CP$ violation in the quark mixing sector, the Kobayashi-Maskawa phase~\cite{Kobayashi:1973fv}, is responsible for a multitude of $CP$-violating phenomena observed in the quark flavor-changing transitions. At the same time, this phase induces electric dipole moments (EDMs) of neutrons and heavy atoms well below current experimental limits. The other source of $CP$ violation, the nonperturbative parameter $\theta$ of quantum chromodynamics (QCD), is largely irrelevant for flavor physics, but tends to induce large EDMs. The nonobservation of EDMs that imply the smallness of theta, $|\theta| \lesssim 10^{-10}$ \cite{Abel:2020pzs,Graner:2016ses}, contrasted with the naive expectation of $\theta \sim { O}(1)$, poses a naturalness problem for the Standard Model, the strong $CP$ problem. 

There are two generic approaches to resolve the strong $CP$ problem. The first approach involves promoting the $\theta$ parameter to a new dynamical field, the QCD axion~\cite{Peccei:1977hh,Weinberg:1977ma, Wilczek:1977pj,Kim:1979if, Shifman:1979if, Dine:1981rt,Zhitnitsky:1980tq}, which symbolically can be represented as 
\begin{equation}
\frac{\theta}{32\pi^2}
G_{\mu\nu}^c {\widetilde G}^{c\mu\nu} ~\to~ \frac12(\partial_\mu a)^2+
\frac{a}{32\pi^2f_a}
G_{\mu\nu}^c {\widetilde G}^{c\mu\nu},
\end{equation}
where $G_{\mu\nu}^c$ is the gluon field strength, ${\widetilde G}^{c\mu\nu}\equiv \frac{1}{2}\varepsilon^{\mu\nu\rho\sigma}G_{\rho\sigma}^c$ with $c$ the adjoint index and $f_a$ is the decay constant of the axion field $a$. The QCD vacuum energy, which for small $\theta$ can be parametrically expressed as 
\begin{equation}
\label{Evac}
     E(\theta)\propto \theta ^2 m_q  \Lambda_{\rm QCD}^3 ~\to ~V(a)= \frac12m_a^2 a^2,
\end{equation}
can be made to dynamically relax to the minimum of the 
potential $V(a)$. In this expression, $\Lambda_{\rm QCD}$ is the nonperturbative scale of the strong interactions, and $m_q$ is the light quark mass. As a result, any initial  
value of $\theta = a/f_a$ will relax to the minimum of the axion potential. In the absence of additional sources of $CP$ violation, this minimum is exactly at $\theta =0$, as in Eq.(\ref{Evac}). Therefore, the neutron EDM that scales as
\begin{equation}
    d_n \propto \frac{m_q\theta}{ \Lambda_{\rm QCD}^2}~,
\end{equation}
is also relaxed to zero. 

Consider now additional sources of $CP$ violation placed at some new physics scale $\Lambda_{\rm CP}$ that we will assume to be larger than the electroweak scale (for example, this could be due to supersymmetric theories with large $CP$-violating phases). Integrating out the new physics at this scale will, in general, result in a number of generic consequences:
\begin{enumerate}
    \item The theta parameter may receive additive corrections to its value, $\theta \to \theta + \theta_{rad}$. Since $G\widetilde G$ is a dimension four operator, $\theta_{rad}$ can depend only on the ratio of scales, and therefore has $\Lambda_{\rm CP}^0$ scaling. Potentially, this can be a large correction, but the axion mechanism will remove the theta term together with $\theta_{rad}$. 
    
    \item $CP$-violating new physics will generically induce higher-dimensional $CP$-odd operators, of which the most relevant are dimension six operators, ${\cal O}_6$ that are suppressed by the square of the new physics scale, 
    and the resulting EDMs will have scaling 
    $d_n({\cal O}_6) \propto \Lambda_{\rm QCD}/\Lambda_{\rm CP}^2$ (or $m_q/\Lambda_{\rm CP}^2$, depending on the chiral properties of ${\cal O}_6$).

    \item In the presence of higher-dimensional $CP$-odd new physics operators, the axion potential minimum shifts away from zero inducing a low-energy value of theta, $\theta_{\rm ind} \propto \Lambda_{\rm QCD}^2/\Lambda_{\rm CP}^2$. This leads to an additional $\theta$-induced contribution to $d_n$ that has, for example, a comparable
    $m_q/\Lambda_{\rm CP}^2$ scaling~\cite{Bigi:1990kz,Pospelov:2000bw,Pospelov:2005pr}. 
    
\end{enumerate}

An important conclusion can be drawn from these observations: the QCD axion mechanism ensures that for sufficiently large $\Lambda_{\rm CP}$, the observable EDMs can be made small and indeed within current bounds for $\Lambda_{\rm CP} \gtrsim 100$\,TeV, one can allow for an arbitrarily large amount of (strong) $CP$ violation above these scales. In this sense, the axion mechanism allows for a proper decoupling of new physics contributions to EDMs. 

The second class of models does not introduce an axion, and instead appeals to symmetry arguments that help to argue why $\theta$ is zero or small.  Historically, models with an exact $CP$ symmetry or exact parity that is spontaneously broken at some UV scale, have been argued to give a viable solution to the strong $CP$ problem (see Refs. \cite{Nelson:1983zb,Barr:1984qx,Babu:1988mw,Babu:1989rb,Mohapatra:1995xd,Kuchimanchi:1995rp,Holdom:1999ny,Hiller:2001qg} for a representative set of ideas). Models based on mirror symmetries have also been used to implement this approach~\cite{Berezhiani:2000gh, Hook:2014cda}.
The most important feature of these models is the absence of a dynamical axion and the 
sensitivity of EDM observables to the value of $\theta$ generated at a UV scale. For example, the spontaneous breaking of $CP$ symmetry may also result in complex quark Yukawa couplings that feed into $\theta_{rad}$ (a representative set of calculations can be found in Refs.~\cite{Ellis:1978hq,Khriplovich:1985jr,Pospelov:1996be,Frampton:1996vxg, deVries:2021pzl}).  Since the $\theta$ term has $\Lambda_{\rm CP}^0$ scaling, this nondecoupling means that all possible sources of $CP$ breaking have to be ``controlled" to very high scales. 

\begin{figure}
    \centering
  {
\begin{tikzpicture}[scale=0.6]
    \shade[top color=black!20,bottom color=black!60]  (3,3.5) rectangle (-5,2);
   \draw [ very thick](-5,2) -- (3,2);\node at (4,2){${M}_{P}$}; 
    \draw (-5,0.5) -- (3,0.5);\node at (4,0.5){${\Lambda}_{CP}$
    }; 
   \draw [dashed](-5,-1) -- (3,-1); \node at (4,-1){$f_a$}; 
    \draw (-5,-2.5) -- (3,-2.5); \node at (4,-2.5){${\Lambda}_{\rm SI}$};  
 \draw (-5,-4) -- (3,-4); \node at (4,-4){$v$
 }; 
     \draw [very thick](-5,-5.5) -- (3,-5.5); \node at (4,-5.5){$\quad{\Lambda}_{\rm QCD}$}; 
     \shade[top color=black!40,bottom color=black!10]  (3,-5.5) rectangle (-5,-6.5) ;
\end{tikzpicture}}

    \caption{Schematic diagram of the different scales referred to in the text. The scale of $CP$ violation, $\Lambda_{\rm CP}$ due to dimension six operators is a UV scale near the Planck scale, $M_P$, and $\Lambda_{\rm SI}
    $ is the small-instanton scale (assumed to be above the electroweak scale, $v$ and QCD strong coupling scale, ${\Lambda}_{\rm QCD}$) where new dynamics enhances the axion mass. The PQ symmetry breaking scale     $f_a$ is assumed to be an independent parameter that can either be above (as shown in the figure) or below the scale $\Lambda_{\rm SI}$. 
    }
    \label{fig:scales}
\end{figure}
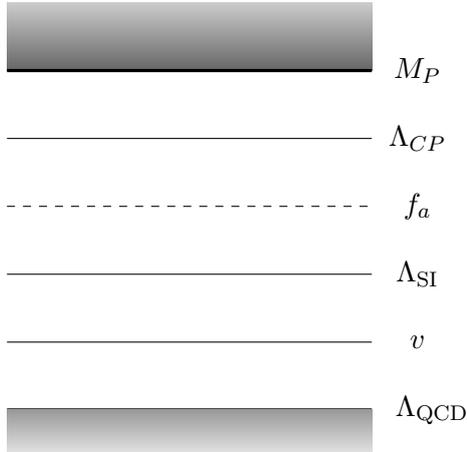

Recently, there has been renewed interest in models that solve the strong $CP$ problem which occupy an intermediate niche between the QCD axion solution and solutions based on discrete symmetries. In this class of models there is still a dynamical axion field and Peccei-Quinn symmetry at a high scale, but the axion mass is now enhanced compared to (\ref{Evac}) by additional dynamical mechanisms at the small-instanton scale $\Lambda_{\rm SI}$. By small instantons we refer to instantons whose size $1/\Lambda_{\rm SI}$ is smaller than the inverse electroweak scale (see Figure~\ref{fig:scales}).
For example, extending the strong gauge interactions and the corresponding axion to a larger group where the non-QCD partners confine at a much larger scale $\Lambda'_{\rm QCD}$ (identified with $\Lambda_{\rm SI}$)
can lead to a significant parametric increase in the axion mass provided $\Lambda'_{\rm QCD} \gg \Lambda_{\rm QCD}$
\cite{Dimopoulos:1979pp,Rubakov:1997vp,Fukuda:2015ana, Gherghetta:2016fhp,Gherghetta:2020ofz}. Similarly, an axion ``portal" between QCD and a mirror QCD with the 
alignment of $\theta$ and $\theta'$ can also result in a heavier axion for $\Lambda'_{\rm QCD} \gg \Lambda_{\rm QCD}$ \cite{Hook:2014cda,Hook:2019qoh}.
Alternatively, if the QCD coupling running is modified to become strong above the TeV scale, the QCD axion mass would receive new contributions from ``small"-size instantons~\cite{Holdom:1982ex, Holdom:1985vx,Dine:1986bg, Flynn:1987rs,  Agrawal:2017ksf, Csaki:2019vte}.
This naturally occurs in models where at some UV scale, QCD propagates in five dimensions~\cite{Gherghetta:2020keg,Gherghetta:2021jnn}.
These models which significantly enhance the axion mass compared to the minimal QCD axion models have a distinctively different phenomenology.
Indeed, given the conventional axion mass range $10^{-6}-10^{-3}$ eV, the enhancement mechanisms imply heavy axions could be in the 100 MeV range or above.
These heavier axions avoid most of the astrophysical bounds, and make the axion amenable to searches at beam dump and collider experiments \cite{Agrawal:2017ksf,Hook:2019qoh}. Moreover, such heavy axions will be less susceptible to possible distortions of the axion potential by the imperfections of the Peccei-Quinn global symmetry.

Besides the enhanced axion mass it is therefore also interesting to consider whether EDM observables could be enhanced in these models. 
In this paper we investigate heavy axion models in the presence of additional sources of $CP$ violation, which are parametrized as higher-dimensional operators that arise from SM fields {and are not related to Planck scale gravitational corrections associated with the axion quality problem.}
The central question we would like to address is whether there is a similar decoupling as for the standard QCD axion, where all observables from, for example, dimension six operators,
scale as $\Lambda_{\rm QCD}^2/\Lambda_{\rm CP}^2$, or if there is an enhancement of $CP$ violation mediated by the induced $\theta$ which is similar to models attempting to solve the strong $CP$ problem using exact parity or $CP$ symmetries.

To answer this question we compute the topological susceptibility and mixed correlators
in heavy axion models that arise from two sources of $CP$ violation: the dimension six Weinberg gluonic operator and a $CP$-odd four-fermion operator. Such $CP$-odd operators induce a linear term in $\theta$ (or equivalently $a$) in the axion potential leading to a shift $\theta_{\rm ind}$ in the potential minimum.  {Similar contributions were proposed in \cite{Holdom:1985vx}, and were estimated on dimensional grounds for fermionic and scalar operators in \cite{Dine:1986bg, Flynn:1987rs, Kitano:2021fdl}.

Instead of relying on dimensional analysis} our computation employs a simple, noninteracting instanton (or anti-instanton) background that ignores strong coupling effects, where we are able to extract qualitative results which show that the induced theta, $\theta_{\rm ind} \propto \Lambda_{\rm SI}^2/\Lambda_{\rm CP}^2$.

This induced shift is qualitatively different from the usual QCD axion scenario and solutions based on exact discrete symmetries due to the presence of the new scale
$\Lambda_{\rm SI}$.
While there is still decoupling in the $\Lambda_{\rm CP} \to \infty$ limit, our results show that the induced $\theta$ can enhance the magnitude of observable EDMs, even to the point that if $\Lambda_{\rm SI}^2/\Lambda_{\rm CP}^2$ is too large, the strong $CP$ problem will reappear. Thus, models with a dynamically enhanced axion mass are subject to bounds depending on the amount of $CP$ violation that is present at energy scales that may significantly exceed 100 TeV. Interestingly, the enhanced EDMs are potentially observable in future EDM experiments.
This paper is organized as follows: 
in Section \ref{sec:InstCF}, we investigate vacuum correlators in an instanton (or anti-instanton) background with different sources of $CP$ violation that shift the axion potential minimum. In Section \ref{sec3}, we consider different heavy QCD axion models with small instantons, deriving the resulting size of the induced $\theta$
and subsequent constraints on the $CP$-violating scale, $\Lambda_{\rm CP}$. We reach our conclusions in Section \ref{sec4}. 

\section{Instanton Correlation Functions}
\label{sec:InstCF}

We begin with briefly reviewing QCD dynamics and the instanton solution that will be used to compute various instanton correlation functions.
The pure Yang-Mills part of the QCD Lagrangian is given by
\begin{equation}\label{eq:QCD_Lag}
    {\cal L}_{\rm QCD} = -\frac{1}{4 g^2} G_{\mu\nu}^a G^{a\mu\nu} + \frac{\theta}{32\pi^2} G_{\mu\nu}^a {\widetilde G}^{a\mu\nu}~,
\end{equation}
where $g$ is the QCD gauge coupling, $\theta$ is the QCD vacuum angle 
and $a=1,\dots, 8$ labels the gauge adjoint representation.
The BPST instanton solution~\cite{Belavin:1975fg} is given by
\begin{equation}
  A_\mu^{a}(x) = \frac{2\eta_{\mu\nu}^a(x-x_0)_\nu}{(x-x_0)^2+\rho^2}~,
          \label{eq:4Dinstanton}
\end{equation}
where the instanton is located at $x_0$ and has a size $\rho$.
The $\eta^a_{\mu\nu}$ denote the group-theoretic 't Hooft $\eta$ symbols~\cite{tHooft:1976snw}. The topological charge is defined to be
\begin{equation}
          Q=\frac{1}{32\pi^2} \int d^4 x\, G_{\mu\nu}^a {\widetilde G}^{a\mu\nu}~,
\end{equation}
where $Q=1$ for the one instanton solution \eqref{eq:4Dinstanton}. We will next compute correlation functions in the instanton (or anti-instanton) background \eqref{eq:4Dinstanton} that will be useful in obtaining contributions to EDM observables such as the neutron EDM.

\subsection{Topological susceptibility}
The vacuum-to-vacuum amplitude in QCD can be written as
\begin{equation}\label{eq:qcd_vac_amp}
   \left\langle 0| 0 \right\rangle
   =\sum_Q\int {\cal D} A^{(Q)}_\mu~ e^{-S_E}~,
\end{equation}
where the Euclidean action for \eqref{eq:QCD_Lag} in an instanton background of charge $Q$ \cite{Jackiw1977ConformalConfigurations} is given by
\begin{equation}\label{eq:QinstantonEuclidean}
    S_{E} = \frac{8\pi^2}{g^2}|Q| +iQ\theta~.
\end{equation}
The topological susceptibility is then introduced as \cite{Witten:1979vv, Shifman:1979if, Bigi:1990kz}
\begin{equation}
   \chi(0)=-i\, \lim_{k\rightarrow 0} \int d^4x \, e^{ikx} \left\langle 0 \left|T\left\{\frac{1}{32\pi^2} G{\widetilde G}(x), \frac{1}{32\pi^2} G{\widetilde G}(0)\right\}\right|0\right\rangle~,
    \label{eq:topsusc}
\end{equation}

where $G{\widetilde G}$ is shorthand notation for $G_{\mu\nu}^a{\widetilde G}^{a\mu\nu}$.

Since the amplitude in the $|Q|>1$ instanton background becomes more exponentially suppressed, only the $Q=\pm 1$ configurations dominate the path integral. Henceforth, we refer to $S_E$ in \eqref{eq:QinstantonEuclidean} only for $|Q|=1$.  In the instanton background \eqref{eq:4Dinstanton} we then obtain the two-point correlator 
\begin{eqnarray}
&& \left\langle 0\left| T\left\{G{\widetilde G}(x), G{\widetilde G}(0)\right\}\right| 0 \right\rangle_{ Q =+ 1} \nonumber \\
 &=& \int {\cal D} A_\mu~ G{\widetilde G}(x) G{\widetilde G}(0)~e^{-\frac{8\pi^2}{g^2_0}}~, \label{eq:instantonint0}\\
 &=&\int d^4 x_0\, \frac{d\rho}{\rho^5}\, C[N] \left( \frac{8\pi^2}{g^2(1/\rho)}\right)^{2N} e^{-\frac{8\pi^2}{g^2(1/\rho)}}
 \frac{192 \rho^4}{((x-x_0)^2+\rho^2)^4} \frac{192 \rho^4}{(x_0^2+\rho^2)^4}~,
 \label{eq:instantonint}
\end{eqnarray}
where the running coupling $g(1/\rho)$ encodes corrections from the quantum fluctuations. In \eqref{eq:instantonint0} we have replaced the path integral over the fluctuation $A_\mu$ with an integration in \eqref{eq:instantonint} over the collective coordinates (see Ref.~\cite{tHooft:1976snw}) where, assuming an $SU(N)$ gauge group,\footnote{In principle, we should also include the normalized Haar measure of the group, as computed in \cite{Cordes:1985um, Csaki:2019vte}. We will omit this measure since its value is simply one for \eqref{eq:instantonint} (or an ${ O}(1)$ number in more generic cases), and therefore our qualitative results remain unchanged.} 
the coefficient
\begin{equation}
    C[N] = \frac{C_1\,e^{-C_2 N}}{(N-1)!(N-2)!} ~,
    \label{eq:Cdef}
\end{equation}
and  $C_1,C_2$ are order one constants ($C_1=0.466, C_2=1.679$ using Pauli-Villars regularization~\cite{Vainshtein:1981wh}). The gauge coupling running is given by
\begin{equation}
       \frac{8\pi^2}{g^2(1/\rho)}= \frac{8\pi^2}{g_0^2}- b_0 \log(M_{UV} \rho)~,
       \label{eq:gaugecoupling}
\end{equation}
where $b_0=4N-N/3=11N/3$ is the pure $SU(N)$ Yang-Mills $\beta$-function coefficient and $g_0=g(M_{UV})$ with UV cutoff $M_{UV}$. 

{ 
In principle, we could consider an ensemble of instantons and anti-instantons~\cite{Callan:1977gz,Diakonov:1983hh,Shuryak:1988rf} to compute correlation functions. However, the qualitative aspects of such an ensemble can be simply captured by one instanton and one anti-instanton~\cite{Diakonov:1985eg, Diakonov:1995qy}, where the (anti-)instantons are assumed to be noninteracting with each other {and can be justified in the weak coupling regime.}
Thus, we will compute correlation functions by adding the contribution from an instanton background to that in an anti-instanton background.}
The total contribution to the topological susceptibility, obtained by performing the $x$ integration first that arises from \eqref{eq:topsusc}, followed by the $x_0$ integration in \eqref{eq:instantonint}, is then given by 
\begin{equation} \label{eq:topsuscint}
     \chi(0)=  -2i\int \frac{d\rho}{\rho^5} C[N] \left( \frac{8\pi^2}{g^2(1/\rho)}\right)^{2N} e^{-\frac{8\pi^2}{g^2(1/\rho)}}~.
\end{equation} 
Assuming an asymptotically free theory, the integral in \eqref{eq:topsuscint} is divergent for large instantons but can be evaluated with a IR cutoff $\rho_{IR}$ on the instanton size.
Assuming $N=3$ with $\rho_{IR}=1/\Lambda_{\rm QCD}$ we obtain $\chi(0) \propto \Lambda_{\rm QCD}^4$.

\subsubsection{Fermion contributions}

The introduction of fermions modifies the path integral and the collective coordinate integration.
In the massless fermion limit, the pure vacuum-to-vacuum transition amplitude is zero. Instead, the instanton now causes transitions from left-handed to right-handed fermions violating the $U(1)$ chiral symmetry so that, for example, $\langle 0| \bar\psi_{Ri} \psi_{Li}|0\rangle \neq 0$. Thus, instantons only contribute to correlation functions in which each fermion flavor and chirality appears at least once.

The effect of massless fermions is usually formulated as an ``effective" Lagrangian~\cite{tHooft:1976snw, Vainshtein:1981wh}
\begin{eqnarray}
        \int d^4 x_0\, {\cal L}_f = \int d^4 x_0\, \frac{d\rho}{\rho^5}\, C[N] e^{0.292 N_f}\left( \frac{8\pi^2}{g^2(1/\rho)}\right)^{2N}  
        e^{-S_E}
        \rho^{3N_f} {\rm det} \left[\bar\psi_{R}(x_0) \psi_{L}(x_0)\right] + h.c.,\nonumber\\
        \label{eq:fermioneffL}
\end{eqnarray}
where the determinant is taken over the $N_f$ fermion flavors, and $\psi_{L,R}^\alpha(x_0)$ are  the fermion zero modes. The constant $e^{0.292 N_f}$ assumes Pauli-Villars regularization and the gauge coupling running \eqref{eq:gaugecoupling} now includes the fermion contributions $b_0 \rightarrow b_0-2/3 N_f$. 

Note that because of the explicit appearance of the fermion zero modes $\psi_{L,R}(x_0)$ in \eqref{eq:fermioneffL}, there is only a contribution to the axion potential if the external fermion zero mode legs are closed. There are two ways this can occur. The first way is to assume that the fermions have an explicit mass $m_f$ (corresponding to a nonzero Higgs vacuum expectation value (VEV), $v\approx 246$~GeV) that connects left- and right-handed fermion fields. The determinant in the effective action then gives a contribution $\propto (\rho\, m_f)^{N_f}$ for $N_f$ fermion flavors. This is the case for the usual contributions from ``large" instantons with $\rho\sim \rho_{IR}=1/\Lambda_{\rm QCD}$ and $m_f \lesssim  \Lambda_{\rm QCD}$. However, since we are interested in ``small" instantons corresponding to instanton sizes ($\sim 1/\Lambda_{\rm SI}$) much smaller than the inverse of the electroweak scale, a second possibility is to close the external fermion zero-mode legs in \eqref{eq:fermioneffL} with $N_f/2$ Higgs bosons. This contribution will be proportional to the product of Yukawa couplings (times a loop factor) and is larger than the Higgs VEV contribution that now scales as $\sim \left({m_f}/{\Lambda_{\rm SI}}\right)^{N_f}$  (assuming $\Lambda_{\rm SI}\gg v$).
Instead of proceeding with the 't Hooft determinant operator in the effective Lagrangian \eqref{eq:fermioneffL} we will follow the approach taken in Refs.~\cite{Flynn:1987rs, Csaki:2019vte} and directly compute the vacuum-to-vacuum amplitude by including the Higgs-fermion Yukawa interaction in the path integral. 

Consider a Higgs field $H$ which couples to $N_f$ flavors of massless fermions with the following Euclidean action
\begin{equation}
      S_H=S_H^{(0)} -i \int d^4 x \sum_{i=1}^{N_f} \frac{y_i}{\sqrt{2}} H(x) \bar\psi_i(x) \psi_i(x)~,
\end{equation}
where $S_H^{(0)}$ is the quadratic (free) part of the Higgs action and $y_i$ are the Yukawa couplings. The Yukawa couplings, or equivalently the fermion masses, have been redefined to be real with their phase included in $\bar\theta=\theta+\text{Arg~Det}M_q$, where $M_q$  is the quark mass matrix. The vacuum-to-vacuum amplitude now takes the 
form
\begin{eqnarray}
      \langle 0|0\rangle_{\Delta Q=1} &=& \int d^4 x_0\, \frac{d\rho}{\rho^5}\, C[N] \left( \frac{8\pi^2}{g^2(1/\rho)}\right)^{2N} e^{-S_E}\nonumber\\
      &&\qquad\times \int {\cal D H}\,e^{-S_H^{(0)}}\,{\cal D \psi}{\cal D \bar\psi}\, e^{~-S_\psi^{(0)}+i \int d^4 x \sum_{i=1}^{N_f} \frac{y_i}{\sqrt{2}} H(x) \bar\psi_i(x) \psi_i(x)}~,\nonumber\\
    &=&  \int d^4 x_0\, \frac{d\rho}{\rho^5}\, C[N] e^{0.292 N_f}\left( \frac{8\pi^2}{g^2(1/\rho)}\right)^{2N} e^{-S_E} (N_f-1)!! \left(\prod_{i=1}^{N_f}\frac{y_i \rho}{\sqrt{2}}\right) ~{\cal I}^{N_f/2}~,\nonumber\\
    \label{eq:vacampF}
\end{eqnarray}
where the action $S_E$ is defined in \eqref{eq:QinstantonEuclidean} with $\theta\rightarrow\Bar{\theta}$. The first line in \eqref{eq:vacampF} shows the collective coordinate integration arising from the gauge field part of the path integral and the second line contains the Higgs and massless fermion contributions to the path integral
with $S_\psi^{(0)}$ the quadratic (free) part of the fermion action.
Integrating over the fermionic fields introduces the factor $e^{0.292 N_f}$ 
and the running gauge coupling now contains fermionic contributions via $b_0\rightarrow b_0 - 2/3 N_f$. Finally, the path integral over the Higgs field gives a nonzero contribution to the amplitude provided all Higgs fields are contracted where 
$(N_f-1)!!$ is the number of Higgs contractions and the quantity ${\cal I}$ is given by \cite{Flynn:1987rs,Csaki:2019vte}

\begin{eqnarray}
   {\cal I} &=&-\int d^4 x_1 \int d^4 x_2\, \bar\psi_i^{(0)}(x_1)\psi_i^{(0)}(x_1)\bar\psi_j^{(0)}(x_2)\psi_j^{(0)}(x_2)\Delta_H(x_1-x_2)~,\nonumber\\
    &=& \frac{\rho^4}{4\pi^8} \int d^4 x_1 \int d^4 x_2\, \int d^4 k\,\frac{1}{k^2+m_H^2}\, \frac{e^{-i k(x_1-x_0)}}{((x_1-x_0)^2+\rho^2)^3}\frac{e^{i k(x_2-x_0)}}{((x_2-x_0)^2+\rho^2)^3}~,\nonumber\\
   &\approx&\begin{cases} \frac{1}{12\pi^2\rho^2}\qquad m_H\rho \ll 1~,\\
   \frac{1}{5\pi^2m_H^2\rho^4}\qquad m_H\rho \gg 1~.
   \end{cases}
   \label{eq:Iintegral}
\end{eqnarray}
In the second line of \eqref{eq:Iintegral} we have substituted for
the scalar Feynman propagator $\Delta_H(x_1-x_2)$ and the fermions have been replaced with their respective zero mode expressions given in \cite{Vainshtein:1981wh}. Note that for an instanton background we have two zero modes $\bar{\psi}_{i,L}^{(0)},{\psi}_{j,R}^{(0)}$ (and $\bar{\psi}_{i,R}^{(0)}, {\psi}_{j,L}^{(0)}$ in an anti-instanton background) where the subscripts $L,~R$, which are suppressed hereon, denote left- and right-handed fields, respectively. Thus, combining \eqref{eq:Iintegral} and \eqref{eq:vacampF} gives the final expression (assuming $m_H \rho \ll 1$)
\begin{equation}\label{eq:fermionCont}
      \langle 0|0\rangle_{\Delta Q=1} =  \int d^4 x_0\, \frac{d\rho}{\rho^5}\, C_f[N] \left( \frac{8\pi^2}{g^2(1/\rho)}\right)^{2N} e^{-S_E} ~,
\end{equation}
with $S_E$ defined in \eqref{eq:QinstantonEuclidean} (assuming $\theta\rightarrow \bar\theta$), and 
\begin{equation}
       C_f[N]\equiv (N_f-1)!! \left(\frac{2}{3}\right)^{N_f/2} \left(\prod_{i=1}^{N_f} \frac{y_i}{4\pi} \right) e^{0.292 N_f} C[N]~.
       \label{eq:Cfdef}
\end{equation}
The expression \eqref{eq:fermionCont} shows how the instanton density in the vacuum-to-vacuum amplitude is modified in the presence of massless fermions and a Higgs-fermion Yukawa interaction. As expected, the amplitude vanishes if any Yukawa coupling is zero. Thus, the topological susceptibility \eqref{eq:topsuscint} 
in the presence of massless fermions is obtained by the substitutions $C[N]\rightarrow C_f[N]$, $\theta\rightarrow \bar\theta$ and $b_0\rightarrow b_0-2/3 N_f$. 

In the case of ``large" instantons associated with the scale $1/\Lambda_{\rm QCD}$, the expression for the vacuum-to-vacuum amplitude differs from \eqref{eq:fermionCont}. As already mentioned, each light fermion ($m_f\lesssim \Lambda_{\rm QCD}$) introduces an $e^{0.292 }\rho\, m_f$ factor
\footnote{In QCD, $\chi(0)\propto m_f$, whereas the $\chi(0)$ resulting from \eqref{eq:QCD:integral} $\propto m_f^{N_L}$. The difference can be understood in terms of instanton-(anti-)instanton interactions-
either via mixing between the fermion zero modes of the instanton with those of the anti-instanton~\cite{Diakonov:1984vw}, or using 't Hooft vertices with fermion legs joined between an instanton and anti-instanton~\cite{Diakonov:1983hh}.}.
This can be seen via the first line in \eqref{eq:Iintegral} where $\Delta_H$ can be replaced by $v^2$, which just gives ${\cal I}=v^2$, and hence:
\begin{eqnarray}
      \langle 0|0\rangle_{\Delta Q=1} =  \int d^4 x_0\, \frac{d\rho}{\rho^5}C[N] \left( \frac{8\pi^2}{g^2(1/\rho)}\right)^{2N} e^{-S_E}
      \left(\prod_{i=1}^{N_L} \rho\, m_i \right)\,e^{0.292 N_L},\label{eq:QCD:integral}
\end{eqnarray}
where the product runs only over $N_L$ light fermions
and $m_i=y_i\,v/\sqrt{2}$.

\subsection{Weinberg gluonic operator}
\label{sec:WeinOp}
The Weinberg operator is a purely gluonic, $CP$ odd, dimension six term given by ${\cal O}_W=GG{\widetilde G}$~\cite{Weinberg:1989dx} that leads to the Lagrangian term
\begin{equation}
    {\cal L} \supset \frac{1}{\Lambda_W^2}GG{\widetilde G}~,
    \label{eq:WLag}
\end{equation}
where $\Lambda_W$ is an effective UV scale. The operator \eqref{eq:WLag} can induce a shift in the axion potential minimum, which can be computed by 
considering the mixed correlator~\cite{Pospelov:2005pr, Weiss:2021kpt}
\begin{eqnarray}
\label{eq:Wsusc}
 \chi_{W}(0)&=&-i\,
\lim_{k\rightarrow 0} \int d^4x \, e^{ikx}
\left\langle 0\left| T\left\{\frac{1}{32\pi^2}G{\widetilde G}(x), \frac{1}{\Lambda_W^2} GG{\widetilde G}(0)\right\}\right|0 \right\rangle~.
\end{eqnarray}
In the instanton background \eqref{eq:4Dinstanton} we obtain 
\begin{equation}
    {\cal O}_W= 
    f_{abc} G_{\mu\kappa}^a G_{\kappa\nu}^b{\widetilde G}^{c\nu\mu}(x) = 
    -\frac{1536 \rho^6}{((x-x_0)^2+\rho^2)^6}~,
     \label{eq:instWeinberg}
\end{equation}
where $f_{abc}$ are the structure constants. Note that for an $SU(N)$ gauge group, the $SU(2)$ instanton solution is embedded in the top left corner of the $N\times N$ matrix of $SU(N)$ generators. Thus, the sum in \eqref{eq:instWeinberg} only gives nonzero contributions for $a,b,c=1,2,3$. 
Furthermore,
 \begin{eqnarray}
&& \left\langle 0\left|  T\left\{G{\widetilde G}(x), GG{\widetilde G}(0)\right\}\right|0\right\rangle_{ Q=+1} \nonumber \\
 &=& \int {\cal D} A_\mu~ G{\widetilde G}(x) GG{\widetilde G}(0)~e^{-\frac{8\pi^2}{g^2_0}}~,\nonumber\\
 &=&\int d^4 x_0\, \frac{d\rho}{\rho^5}\, C[N] \left( \frac{8\pi^2}{g^2(1/\rho)}\right)^{2N} e^{-\frac{8\pi^2}{g^2(1/\rho)}}
 \frac{192 \rho^4}{((x-x_0)^2+\rho^2)^4} \frac{-1536 \rho^6}{(x_0^2+\rho^2)^6}~.
 \label{eq:Wcorrelator}
\end{eqnarray}
Again performing the integrals first over $x$ and then $x_0$ gives
\begin{equation}
    \chi_{W}(0)=2i\frac{384\pi^2}{5\Lambda_W^2}\int \frac{d\rho}{\rho^7} C[N] \left( \frac{8\pi^2}{g^2(1/\rho)}\right)^{2N} e^{-\frac{8\pi^2}{g^2(1/\rho)}}~,
    \label{eq:Wmixed}
\end{equation}
where we have also included the anti-instanton contribution.

In the presence of fermions, $\chi_{W}(0)$ is obtained by making the substitutions $C[N]\rightarrow C_f[N]$ for small instantons
(or by introducing the factor $(\rho\,m_f)^{N_L}$, as in \eqref{eq:QCD:integral} for large instantons), $\theta\rightarrow \bar\theta$  and $b_0\rightarrow b_0-2/3 N_f$ in the running gauge coupling $g(1/\rho)$. 

\subsection{Four-fermion operators }
Another class of dimension six operators which can affect the axion solution are the four-fermion operators. Such operators are suppressed by an effective mass scale $\Lambda_F$ and given by
 \begin{equation}\label{Fermi_operator}
\mathcal{L}\supset\sum_{ijkl}\frac{\lambda_{ijkl}}{\Lambda^2_F}\Bar{\psi}_i\psi_j\Bar{\psi}_k\psi_l~,
\end{equation}
where $\lambda_{ijkl}$ are complex coefficients with flavor indices $i,j,k,l$. Note that the spinor and electroweak structure has been suppressed in \eqref{Fermi_operator}, although it is straightforward to incorporate these details. Of particular interest is the spinor structure of \eqref{Fermi_operator} resulting in $CP$ violation. These are operators of the type ${\cal O}_{F,ijkl}=\Bar{\psi}_ii\gamma_5\psi_j\Bar{\psi}_k\psi_l$ which are anti-Hermitian with the corresponding $\lambda_{ijkl}$ purely imaginary.

\begin{figure}[h!]
    \centering
  \centering 
\subfigure[]
{
    \begin{tikzpicture}[scale=0.450]
    \begin{feynhand}
    \setlength{\feynhandblobsize}{8mm}

 \vertex  (h1) at (-1.2941/1.4,4.82963/1.4);  \vertex  (h2) at (-4.82963/1.4,1.2941/1.4); \vertex  (h3) at (-3.53553/1.4,-3.53553/1.4); \vertex  (h4) at (1.2941/1.4,-4.82963/1.4);  \vertex  [dot] (f4) at (4.95722/1.4,0.652631/1.4) {};
    \vertex (v1) at (1.29904/2,.75/2);
     \vertex (v2) at (.75/2,1.29904/2);
       \vertex (v3) at (0,1.5/2);
        \vertex (v4) at (-0.75/2,1.29904/2);
        \vertex (v5) at (-1.29904/2,0.75/2);
     \vertex (v6) at (-1.5/2,0);
     \vertex (v7) at (-1.29904/2,-0.75/2);
      \vertex (v8) at (-0.75/2,-1.29904/2);
       \vertex (v9) at (0,-1.5/2);
        \vertex (v10) at (.75/2,-1.29904/2);
        \vertex (v11) at (1.29904/2,-.75/2);
          \vertex (v12) at (1.5/2,0);  
    \propag[fer] (h1) to [quarter left, edge label'=$t$](v3);
     \propag[fer] (v4) to [quarter left](h1);
      \propag[fer] (h2) to [quarter left, edge label'=$b$](v5);
     \propag[fer] (v6) to [quarter left](h2);
      \propag[fer] (h3) to [quarter left, edge label'=$s$](v7);
     \propag[fer] (v8) to [quarter left](h3);
      \propag[fer] (h4) to [quarter left, edge label'=$c$](v9);
     \propag[fer] (v10) to [quarter left](h4);
      \propag[fer] (v1) to [quarter left](f4);
     \propag[fer] (v11) to [half right, looseness=1.4](f4); 
     \propag[fer] (f4) to [half right, looseness=1.4](v2);
      \node at (6/2.3,1.3) {$u$}; 
        \node at (6/2.3,-0.65) {$d$}; 
   
     \propag[fer] (f4) to  [quarter left](v12);
     \propag[sca](h1) to [quarter right, looseness=1.1, edge label'=$H$](h2);
      \propag[sca](h3) to  [quarter right, looseness=1.1, edge label'=$H$](h4);
       \vertex[grayblob] (tv) at (0,0) {};
       \node at (0,0) {{I}};
       \node at (6/1.3,0.65/1.4) {$\mathcal{O}_F$}; 
    \end{feynhand}
    \end{tikzpicture}    }
    \hspace{12mm} \subfigure[]
{
    \begin{tikzpicture}[scale=0.450]
    \begin{feynhand}
    \setlength{\feynhandblobsize}{8mm}
 \vertex   [dot](f1) at (-3.04381/1.4,3.96677/1.4){} ;  \vertex  [dot] (f2) at (-1.91342/1.4,-4.6194/1.4) {};  \vertex [dot]  (f4) at (4.95722/1.4,0.652631/1.4) {};
   \node at (6/1.3,0.65/1.3) {$\mathcal{O}_F$}; 
     \node at (-3.4/1.2,4.5/1.3) {$\mathcal{O}_F$}; 
       \node at (-2/1.2,-5.2/1.3) {$\mathcal{O}_F$}; 
    \vertex (v1) at (1.29904/2,.75/2);
     \vertex (v2) at (.75/2,1.29904/2);
       \vertex (v3) at (0,1.5/2);
        \vertex (v4) at (-0.75/2,1.29904/2);
        \vertex (v5) at (-1.29904/2,0.75/2);
     \vertex (v6) at (-1.5/2,0);
     \vertex (v7) at (-1.29904/2,-0.75/2);
      \vertex (v8) at (-0.75/2,-1.29904/2);
       \vertex (v9) at (0,-1.5/2);
        \vertex (v10) at (.75/2,-1.29904/2);
        \vertex (v11) at (1.29904/2,-.75/2);
          \vertex (v12) at (1.5/2,0);
           \node at (6/2.3,1.3) {$u$}; 
     \node at (-0.9,2.73) {$t$}; 
       \node at (-0.1,-3) {$c$}; 
        \node at (6/2.3,-0.65) {$d$}; 
     \node at (-2.45,01.45) {$b$}; 
       \node at (-1.9,-2.2) {$s$}; 
   \propag[fer] (v5) to [quarter left](f1);
     \propag[fer] (v3) to [half right, looseness=1.4](f1); 
     \propag[fer] (f1) to [half right, looseness=1.4](v6);
     \propag[fer] (f1) to  [quarter left](v4); 
     
     \propag[fer] (v9) to [quarter left](f2);
     \propag[fer] (v7) to [half right, looseness=1.4](f2); 
     \propag[fer] (f2) to [half right, looseness=1.4](v10);
     \propag[fer] (f2) to  [quarter left](v8);
     
      \propag[fer] (v1) to [quarter left](f4);
     \propag[fer] (v11) to [half right, looseness=1.4](f4); 
     \propag[fer] (f4) to [half right, looseness=1.4](v2);
     \propag[fer] (f4) to  [quarter left](v12);
    \vertex[grayblob] (tv) at (0,0) {};
     \node at (0,0) {{I}};
    \end{feynhand}\label{fig:cpv2}
    \end{tikzpicture}
}\caption{The t'Hooft vertex that includes the insertion of four-fermion operators. Fermion legs are closed with one four-fermion operator ${\cal O}_{F}$ and two Higgs-fermion Yukawa interactions (a) and three four-fermion operators ${\cal O}_F$ (b).}
    \label{fig:4fermioncpv}
\end{figure}
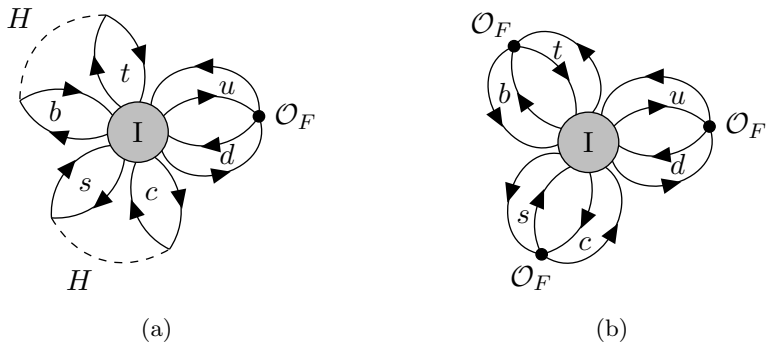

The $CP$-violating effect arising from \eqref{Fermi_operator} can be obtained by including the four-fermion interactions in the path integral \eqref{eq:vacampF}. These operators allow for new ways to close the fermion legs in the 't Hooft vertex, as depicted in Figure~\ref{fig:4fermioncpv}. The largest contribution arises from just one insertion of ${\cal O}_F$, as shown in Figure \ref{fig:4fermioncpv}(a), while more insertions of the four-fermion operator, such as in Figure  \ref{fig:4fermioncpv}(b) are suppressed by powers of $\Lambda_F$.
Similar to the definition \eqref{eq:Wsusc} for $ \chi_W(0)$ we can define a fermion mixed correlator

\begin{eqnarray}
 \chi_{F,ijkl}(0)&=&-i\, \lim_{k\rightarrow 0} \int d^4x \, e^{ikx}
 \left\langle 0\left| T\left\{\frac{1}{32\pi^2}G{\widetilde G}(x),
 \frac{\lambda_{ijkl}}{\Lambda^2_F}{\cal O}_{F,ijkl}(0)\right\}\right|0 \right\rangle~.
 \label{eq:4FermiSuscDef}
 \end{eqnarray}

The only operators contributing to the fermion path integral are those with two pairs of flavor indices ($i=j\neq k=l$ or $i=l\neq k=j$), i.e. ${\cal O}_{F,iijj},~\text{and}~{\cal O}_{F,ijji}$, both of which are hereon generically referred to as ${\cal O}_{F,ij}$ with the corresponding coupling constant $\lambda_{ij}\equiv \lambda_{iijj}$ (or $\lambda_{ijij})$. The explicit expression for such a generic operator $\mathcal{O}_{F,ij}$ can be computed as
\begin{eqnarray}
\chi_{F,ij}(0)&=&-2i\,\int d^4 x_0\, \frac{d\rho}{\rho^5}\, C[N]e^{0.292 N_f}\left( \frac{8\pi^2}{g^2(1/\rho)}\right)^{2N} e^{-\frac{8\pi^2}{g^2(1/\rho)}} \nonumber \\
&\times&\frac{2\lambda_{ij}}{y_iy_j}(N_f-3)!! \left(\prod_{k=1}^{N_f}\frac{y_k \rho }{\sqrt{2}}\right) ~{\cal I}^{N_f/2-1}\frac{1}{\Lambda^2_F}\Bar{\psi}_i^{(0)}i\gamma_5\psi_i^{(0)}\Bar{\psi}_j^{(0)}\psi_j^{(0)}(0)\frac{1}{32\pi^2} \int d^4x\, G{\widetilde G}(x)
~, \nonumber \\
&=&2i~\frac{2(-i\lambda_{ij})}{y_iy_j}\int  \frac{d\rho}{\rho^5}\,\frac{C_f[N]}{N_f-1}  \left( \frac{8\pi^2}{g^2(1/\rho)}\right)^{2N}\frac{12}{5\rho^2\Lambda_F^2}~e^{-\frac{8\pi^2}{g^2(1/\rho)}}~,\label{eq:4FermiIntegral}
\end{eqnarray}
where we have also included the effect of the anti-instanton. The part of ${\cal O}_{F,ij}$ contributing to the path integral in the instanton background is $
i{\psi}^{\dagger}_{L,i}\psi_{R,i}{\psi}^{\dagger}_{L,j}\psi_{R,j} $, while in the anti-instanton background (where $G{\widetilde G}\rightarrow -G{\widetilde G}$) it is $-i{\psi}^{\dagger}_{R,i}\psi_{L,i}{\psi}^{\dagger}_{R,j}\psi_{L,j} $. These two contributions add up\footnote{Instead, for $CP$-even operators of the type $\Bar{\psi}_i\psi_i\Bar{\psi}_j\psi_j$ there is a cancellation between the two contributions since $
{\psi}^{\dagger}_{L,i}\psi_{R,i}{\psi}^{\dagger}_{L,j}\psi_{R,j} $ and ${\psi}^{\dagger}_{R,i}\psi_{L,i}{\psi}^{\dagger}_{R,j}\psi_{L,j} $ both appear with the same sign.} to give the factor of $2i$ in \eqref{eq:4FermiIntegral}. 

The result \eqref{eq:4FermiIntegral} can also be understood in terms of the results \eqref{eq:vacampF} and \eqref{eq:Iintegral} from the fermionic path integral, up to the overall ratio of couplings. If we assume that ${\cal O}_F$ is generated by a heavy scalar of mass $\Lambda_F$, interacting with Standard Model quarks via Yukawa interactions, \eqref{eq:Iintegral} implies a factor of $12\pi^2\rho^2/5\pi^2\Lambda_F^2\rho^4 = 12/5\rho^2 \Lambda_F^2$ relative to the expression \eqref{eq:fermionCont}, which matches the factor inside the integral. The factor $1/({N_f-1})$ arises from having a fewer number of contractions-$({N_f-3})!!$ compared to \eqref{eq:vacampF}, assuming only one insertion of the operator ${\cal O}_{F,ij}.$

Furthermore, notice that $y_i$ and $y_j$ have been explicitly factored out of 
\eqref{eq:4FermiIntegral} to write the result in terms of $C_f[N]$ defined in \eqref{eq:Cfdef}. For $-i\lambda_{ij}\sim 1$, this shows that the effect of the four-fermion operator, being 
$\propto 1/y_iy_j$, is most enhanced for the up and down quarks compared to that from the Weinberg gluonic operator or the second and third generation quarks. However, the four-fermion operator coefficient $\lambda_{ij}$ can be chirally suppressed by Yukawa couplings~\cite{Kitano:2021fdl}. For example, such four-fermion operators with a chiral suppression can arise from the overlap of fermion profiles in extra dimension models~\cite{Bonnefoy:2020llz}. Thus, we will henceforth assume that $-i\lambda_{ij} \propto
 y_i y_j$ so that the effect of the four-fermion operator is similar to that of the Weinberg gluonic operator as well as the contributions from the other generations of quarks.

Assuming $-i\lambda_{ij}=y_iy_j/2$, we then have $N_f(N_f-1)$ contributions of the fermion susceptibility \eqref{eq:4FermiIntegral} for both types of operators ${\cal O}_{F,iijj},~\text{and}~{\cal O}_{F,ijji}$, each.
Thus, for $N_f=6$ we obtain
\begin{eqnarray}
\chi_{F}(0)\equiv 2 N_f(N_f-1)\chi_{F,ij}(0)=2i \frac{144}{5\Lambda_F^2}\int  \frac{d\rho}{\rho^7}\,C_f[N]  \left( \frac{8\pi^2}{g^2(1/\rho)}\right)^{2N}
e^{-\frac{8\pi^2}{g^2(1/\rho)}}~.\label{eq:FermiSusc6}
\end{eqnarray}
 Using \eqref{eq:FermiSusc6} we will place limits on a generic scale $\Lambda_F$ that represents all of these fermion effects.

Finally, note that in supersymmetric theories the operator ${\cal O}_F$ can arise from a dimension-four term in the superpotential~\cite{Pospelov:2005ks}. After integrating out the scalar superpartners this leads to a four-fermion term with
\begin{equation}
    \frac{1}{\Lambda_F^2} \sim \frac{g^2}{16\pi^2}\frac{1}{\Lambda_{\rm UV} m_{\rm SUSY}}~,
\end{equation}
where $\Lambda_{\rm UV}$ is the UV scale of the superpotential term and $m_{\rm SUSY}$ is the supersymmetry-breaking scale of the scalar superpartners. The bounds on $\Lambda_F$ can thus be interpreted as bounds on the scalar superpartner masses.

\section{Induced Theta}\label{sec3}
Using the results in Section~\ref{sec:InstCF} 
we can now obtain an estimate for the shift in the axion potential minimum due to $CP$-odd operators. In the presence of the Weinberg operator the axion potential is modified by a linear term in the axion field
\begin{equation}\label{axion_pot_F}
    V(a) = {\chi_{W}(0)} \left(\frac{a}{f_a}\right) +\frac{1}{2} \chi{(0)} \left(\frac{a}{f_a} \right)^2~,
\end{equation}
where we have promoted the theta angle to the axion field, $\bar{\theta}\rightarrow a/f_a$. This leads to a shift in the potential minimum by an amount
\begin{equation}
  \left \langle \frac{a}{f_a}\right\rangle\equiv \theta_{\rm ind} = -\frac{\chi_{W}(0)}{\chi{(0)}}~.
    \label{eq:indth}
\end{equation}
In the case of four-fermion operators the linear potential term again causes a shift in the potential minimum given by \eqref{eq:indth}, with $\chi_{W}(0)$ replaced by $\chi_{F}(0)$.

The induced $\theta$ then directly contributes to EDM observables such as the neutron EDM where
\begin{equation}
    d_n~ \propto~ \frac{m_q}{ \Lambda_{\rm QCD}^2} |{\theta}_{\rm ind}|~=~
    \frac{m_q}{ \Lambda_{\rm QCD}^2}\frac{\chi_{W,F}(0)}{\chi{(0)}}~.
    \label{dn}
\end{equation}
The experimental limit arising from the neutron EDM gives the constraint
\begin{equation}
|\theta_{\rm ind}| \lesssim 10^{-10}~,
\label{eq:indthlimit}
\end{equation}
which can now be used to obtain constraints on various heavy axion scenarios\footnote{For simplicity, we will present limits that arise from the individual operators ${\cal O}_W$ and ${\cal O}_F$ separately. Our results can be straightforwardly generalized by summing the contributions in \eqref{eq:indth} if both operators are present.}.

\subsection{QCD}
We first consider the effect of dimension six operators in QCD with $N_L$ light fermions (i.e. $m_f\lesssim  \Lambda_{\rm QCD}$). 
The induced $\theta$ \eqref{eq:indth} that arises from including the Weinberg operator is given by
\begin{equation}\theta_{\rm ind}^{\rm QCD}\approx \xi_W\frac{b_0-4+N_L}{b_0-6+N_L} \frac{\Lambda_{\rm QCD}^2}{\Lambda_{W}^2}~,
    \label{eq:QCDindth}
\end{equation}
where $\xi_W=384\pi^2/5$,
 $b_0$ is the $\beta$-function coefficient and the $CP$-violation scale $\Lambda_{\rm CP}$ is identified with $\Lambda_{W}$. Note that in \eqref{eq:QCDindth} the product of all light quark masses cancel and the induced $\theta$ becomes small (or decouples) as $\Lambda_{W}\rightarrow \infty$.
Imposing the constraint \eqref{eq:indthlimit} 
for QCD ($b_0^{\rm QCD}=9$, $N_L=3$ and  $\Lambda_{\rm QCD}\approx 300 ~$MeV),
gives the limit $\Lambda_W\gtrsim 10^6$~GeV on the effective scale of the Weinberg operator.

For the case of the $CP$-odd four-fermion operator, the 't Hooft vertex now has two fewer factors of $\rho\, m_f$ compared to the topological susceptibility resulting from \eqref{eq:QCD:integral}. This gives a bound similar to $\Lambda_W$ when there is no chirality suppression in the four-fermion operator, otherwise the $\Lambda_F$ bound is much weaker. A calculation for $\theta_{\rm ind}$
using the chiral anomaly can be found in \cite{An:2009zh}, which agrees with our estimate of the bound on $\Lambda_F$ within an order of magnitude.

As such, current constraints on the neutron EDM correspond to new $CP$-violating physics at $\sim 10^6$~GeV.
Thus, future neutron EDM experiments can probe new $CP$-violating sources at scales ranging from $\sim 10^6 - 10^9$~GeV, beyond which the SM contribution due to the CKM phase becomes comparable in size.

\subsection{4D Small Instantons}\label{4DInst}
\subsubsection{Product gauge group}
A heavy axion can be generated by extending the QCD gauge group into a product gauge group $SU(3)^k=SU(3)_1\times SU(3)_2\times\dots \times SU(3)_k$ which is spontaneously broken at a scale $\Lambda_{\rm SI}$~\cite{Agrawal:2017ksf,Csaki:2019vte}.
Small instantons at the scale $\Lambda_{\rm SI}$ associated with the product gauge groups lead to this enhancement.
 The SM quarks are assumed to be charged under only $SU(3)_1$. In addition, there are $k$ axions, labeled by $i$, which couple to the $k$ SU(3) $G\widetilde G$ terms with decay constants $f_{a_i}$, eliminating the $k$ theta terms.

At the scale $\Lambda_{\rm SI}$ the QCD gauge coupling $\alpha$ is matched to the $SU(3)_k$ gauge couplings, $\alpha_i$ via the relation
\begin{equation}
    \frac{1}{\alpha(\Lambda_{\rm SI})}=\sum_{i=1}^k \frac{1}{\alpha_i(\Lambda_{\rm SI})}~.
\end{equation}
This relation implies that each individual coupling $\alpha_i$ must be larger than the QCD coupling at the scale $\Lambda_{\rm SI}$. Therefore, the larger couplings $\alpha_i(\Lambda_{\rm SI})$ can make the small instanton effects dominate over the usual QCD large instantons. This effect is most dominant in the limit $k\gg 1$, where the axion masses scale as $m_{a_1} \sim \sqrt{\Pi_f y_f} {\Lambda^2_{\rm SI}}/f_{a_1}$ (with $y_f$ the quark Yukawa couplings) and $m_{a_i} \sim {\Lambda^2_{\rm SI}}/f_{a_i}$ for $i=2\dots, k$, showing that the lightest axion mass ($m_{a_1}$) can remain much heavier that the QCD axion mass for
$\Lambda_{\rm SI}\gg \Lambda_{\rm QCD}$.  

For concreteness, let us consider the case with small $k$, where there is some perturbative control and the instanton (or anti-instanton) background still gives us qualitatively accurate results. Assuming the product gauge group is broken by scalars with a VEV, $v_\phi $,
the effective cutoff for the instanton size then becomes $2\pi v_\phi$,
in contrast to the naive expectation, $\Lambda_{\rm SI}$~\cite{Csaki:2019vte}.
The constraint \eqref{eq:indthlimit} can then be used to obtain limits on the scales associated with the sources of $CP$ violation from the Weinberg and four-fermion operators.
Since the QCD instanton contribution to $\chi_{W,F}(0)$ is suppressed by at least ${\Lambda_{\rm QCD}^2}/{\Lambda_{W,F}^2}$, the small instanton contribution from the UV gauge group dominates and results in
\begin{equation}
\theta_{\rm ind}\approx
 \xi_{W,F}\frac{2}{b_{0,i}-6} \frac{(2\pi v_\phi)^2}{\Lambda_{W,F}^2}\approx
\xi_{W,F}\frac{8\pi^2}{b_{0,i}-6}
\frac{\Lambda_{\rm SI}^2}{\Lambda_{W,F}^2}~,
    \label{eq:SIindth_product}
\end{equation}
where 
$\xi_F=24 N_f/5$, $\xi_W$ is defined under \eqref{eq:QCDindth}
 and we have assumed $\Lambda_{\rm SI}
\approx v_\phi$ in the second expression in \eqref{eq:SIindth_product}.
   The constraint  \eqref{eq:indthlimit} then
 implies $\Lambda_{\rm SI}/\Lambda_{W}\lesssim 10^{-8}$ and $\Lambda_{\rm SI}/\Lambda_{F}\lesssim 10^{-7}$ or $\Lambda_{\rm SI}\lesssim 10^{10}(10^{11})$~GeV for $\Lambda_{W} (\Lambda_F)=M_P$ where $M_P=2.4\times 10^{18}$~GeV is the (reduced) Planck mass\footnote{The difference in these two bounds results from the size of the different prefactors $\xi_{W,F}$, where $\xi_W$ results from the large number of color contractions in \eqref{eq:Wsusc}, while $\xi_F$ arises from the smaller flavor multiplicity of the four-fermion operator \eqref{Fermi_operator}.}, $b_{0,1}=13/2$ and $b_{0,k}=21/2$. For $i=2,\dots, k-1$, the same expression \eqref{eq:SIindth_product} holds with $b_{0,i}=10$, and $v_\phi\rightarrow \sqrt{2} v_\phi$, which does not change the bounds significantly\footnote{It is possible that the axion mass could instead be dominated by QCD large instantons. But
in this case the $CP$ violation arising from  small instantons of the product gauge group gives the much weaker constraint that $\Lambda_{\rm SI}/\Lambda_{W}\lesssim 10^{-8}\times m_{a, \rm QCD}/m_{a_1}$. For instance, assuming $m_{a,\rm QCD}/m_{a_1}=10^3$ implies that $\Lambda_{\rm SI}\lesssim 10^{13}$~GeV for $\Lambda_W=M_P$.}. 
Note that if UV couplings are included in \eqref{eq:WLag} then the effective scale $\Lambda_W$ can be larger than $M_P$. 
 Assuming $f_a>\Lambda_{\rm SI}$, the limits on $\Lambda_{W,F}$ correspond to a maximum possible axion mass enhancement of $\sim 10^7$ for $k=3$ relative to the QCD axion~\cite{Agrawal:2017ksf,Csaki:2019vte}. As such, axion masses $m_a\gtrsim 100$~MeV with $f_a\lesssim 10^7$~GeV~\cite{Cadamuro:2011fd, Jaeckel:2015jla} can be explored in future experimental searches.

{ However, when 
$f_a<\Lambda_{\rm SI}$, we need to UV complete the dimension five axion-$G\widetilde G$ coupling and explain the PQ breaking. This  can be done in a minimal KSVZ-type scenario~\cite{Shifman:1979if, Kim:1979if}, by introducing a single heavy Dirac fermion $\Psi$, with mass $m_\Psi$, charged under the $U(1)_{PQ}$ symmetry, which changes the instanton measure by a factor of $e^{0.292}\rho\, m_\Psi$.
Combining this with the contribution arising from the running of the gauge coupling between $m_\Psi$ and $\Lambda_{\rm SI}$, the topological susceptibility (or any similar correlator) is modified to
\begin{eqnarray}
    \chi(0)&\rightarrow& 
    \chi(0)\,
    \frac{b_0-4}{b_0-11/3}
    \left(\frac{m_\Psi}{\Lambda_{\rm SI}}\right)^{-2/3}\,e^{0.292}\left(\frac{m_\Psi}{\Lambda_{\rm SI}}\right)\approx
    \chi(0)\left(\frac{f_a}{\Lambda_{\rm SI}}\right)^{1/3}~,
    \label{eq:topsussuppress}
\end{eqnarray}
where the Yukawa coupling between $\Psi$ and the PQ scalar is assumed to be order one, i.e. $m_\Psi\approx f_a$. Since $m_a^2\propto \chi(0)$ this suppresses the axion mass enhancement by an amount $(
f_a/\Lambda_{\rm SI})^{1/6}$~\cite{Co:2022aav}.
Thus for the experimentally interesting region of $m_a\gtrsim 100$~MeV and $f_a\lesssim 10^7$~GeV, the axion mass enhancement is reduced by up to a factor of 10 when $f_a<\Lambda_{\rm SI}$.}

A similar result is also obtained for an enlarged color group~\cite{Holdom:1982ex, Holdom:1985vx,Gherghetta:2016fhp} where $\Lambda_{\rm SI}$ is identified with the scale where the enlarged symmetry group is broken and the appropriate $b_0$ is used.
In all these cases, there is again a nondecoupling effect that depends on the ratio $\Lambda_{\rm SI}/\Lambda_{W,F}$.

\subsubsection{Mirror QCD}

A heavy axion can also be obtained by assuming that there exists a $\mathbb{ Z}_2$ mirror copy of QCD that becomes strong at a scale 
$\Lambda^\prime_{\rm QCD}(\equiv\Lambda_{\rm SI})
\gg \Lambda_{\rm QCD}$~\cite{Rubakov:1997vp,Berezhiani:2000gh,Hook:2014cda,Fukuda:2015ana,Hook:2019qoh}. The axion is $\mathbb{Z}_2$ neutral and couples to both QCD and mirror QCD, via the interaction
\begin{align}
    \frac{1}{32\pi^2}\frac{a}{f_a}\varepsilon^{\mu\nu \rho \sigma}\left(G^c_{\mu\nu} G_{\rho \sigma }^c + G_{\mu\nu}^{\prime c} G_{\rho \sigma }^{\prime c}\right),
\end{align}
where $G'_{\mu\nu}$ is the mirror QCD field strength. The axion now receives contributions from the mirror QCD instantons (which are small {in size} relative to those from QCD)
{and gives rise to limits on higher dimensional operators with scales $\Lambda_{W,F}$ involving gluons and fermions in the mirror sector.}

The mirror QCD expression for the induced $\theta$ due to the Weinberg operator can be obtained by substituting $\Lambda_{\rm SI}$ in \eqref{eq:QCDindth}.
This leads to the bounds $\Lambda_{\rm SI}/\Lambda_W \lesssim  10^{-7}$ or $\Lambda_{\rm SI} \lesssim  10^{11}$~GeV for $\Lambda_W = M_P$, assuming the mirror Higgs VEV, $v' \gg \Lambda_{\rm SI}$ such that QCD$^{\prime}$ is a pure Yang-Mills theory at $\Lambda_{\rm SI}$ with $b_0^{\rm QCD'}=11$.
 These bounds for the Weinberg operator do not change appreciably if this assumption is relaxed. 

The induced $\theta$ from the four-fermion operator can be obtained by considering $N_L\geq 2$ light flavors in QCD$^\prime$. Applying the QCD result \eqref{eq:QCD:integral} for QCD$^\prime$ then gives
\begin{eqnarray}
\theta_{\rm ind, F}&\approx& \frac{2 N_L(N_L-1)}{5\pi^2}\frac{b_0-4+N_L}{b_0-8+N_L} \frac{\Lambda_{\rm SI}^4}{v'^2\,\Lambda_{F}^2}
\approx\frac{2 N_L(N_L-1)}{5\pi^2}\frac{b_0-4+N_L}{b_0-8+N_L} \frac{\Lambda_{\rm SI}^2}{\Lambda_{F}^2}~, \label{eq:QCDFindth}
\end{eqnarray} 
where we have taken $v'\approx\Lambda_{\rm SI}$ in the last expression in \eqref{eq:QCDFindth}.
Assuming $b_0=9$ and $N_L=3$, implies $\Lambda_{\rm SI}/\Lambda_F \lesssim  10^{-5}$, or $\Lambda_{\rm SI} \lesssim  10^{13}$~GeV for $\Lambda_{F} = M_P$. Again, the difference in the $\Lambda_{W,F}$ bounds arises from the different color and flavor multiplicity factors.

\subsection{5D Small Instantons}\label{5DInst}

Another way for the QCD coupling to become large at a UV scale and increase the effect of small instantons is to consider a 5D model where QCD gluons propagate in a fifth dimension of size $R$.
The axion can be identified with a UV boundary localized field that couples to QCD via a coupling proportional to $1/f_a$, with { $f_a$ an independent parameter of the theory.} {This allows the decay constant to be either above or below the small instanton scale and allows for more general possibilities.} Above the scale $1/R$ the QCD coupling increases in strength until the coupling becomes strong at the cutoff scale $\Lambda_5$ which is defined by the relation~\cite{Gherghetta:2020keg}
\begin{equation}
\Lambda_5 R=\frac{6\pi\epsilon}{\alpha(1/R)}~,
\end{equation}
where $\alpha=g^2/(4\pi)$ and $\epsilon\leq 1$ is a perturbativity parameter\footnote{{ Note that in the 5D model, small instantons can be made to dominate when perturbativity still holds. This implies that our instanton (or anti-instanton) approximation used for the correlators will give more accurate quantitative results relative to QCD.}}. The small instanton scale can be identified as $\Lambda_{\rm SI}\equiv \Lambda_5$.
The 4D effective action is approximately given 
by~\cite{Gherghetta:2020keg}
\begin{equation}
    S_{\rm eff}\approx \frac{2\pi}{\alpha_s(1/R)}-\frac{R}{\rho}+b_0\ln\frac{R}{\rho}~,
    \label{eff_action}
\end{equation}
where the power-law term $R/\rho$ arises from summing over the 5D Kaluza-Klein gluons. Thus, small instantons of size $1/\Lambda_{\rm SI}\lesssim \rho \lesssim R$ can now reduce the effective action and contribute greatly to the path integral. 

\begin{figure}[h!]
   \centering
      \includegraphics[width=0.8\textwidth]{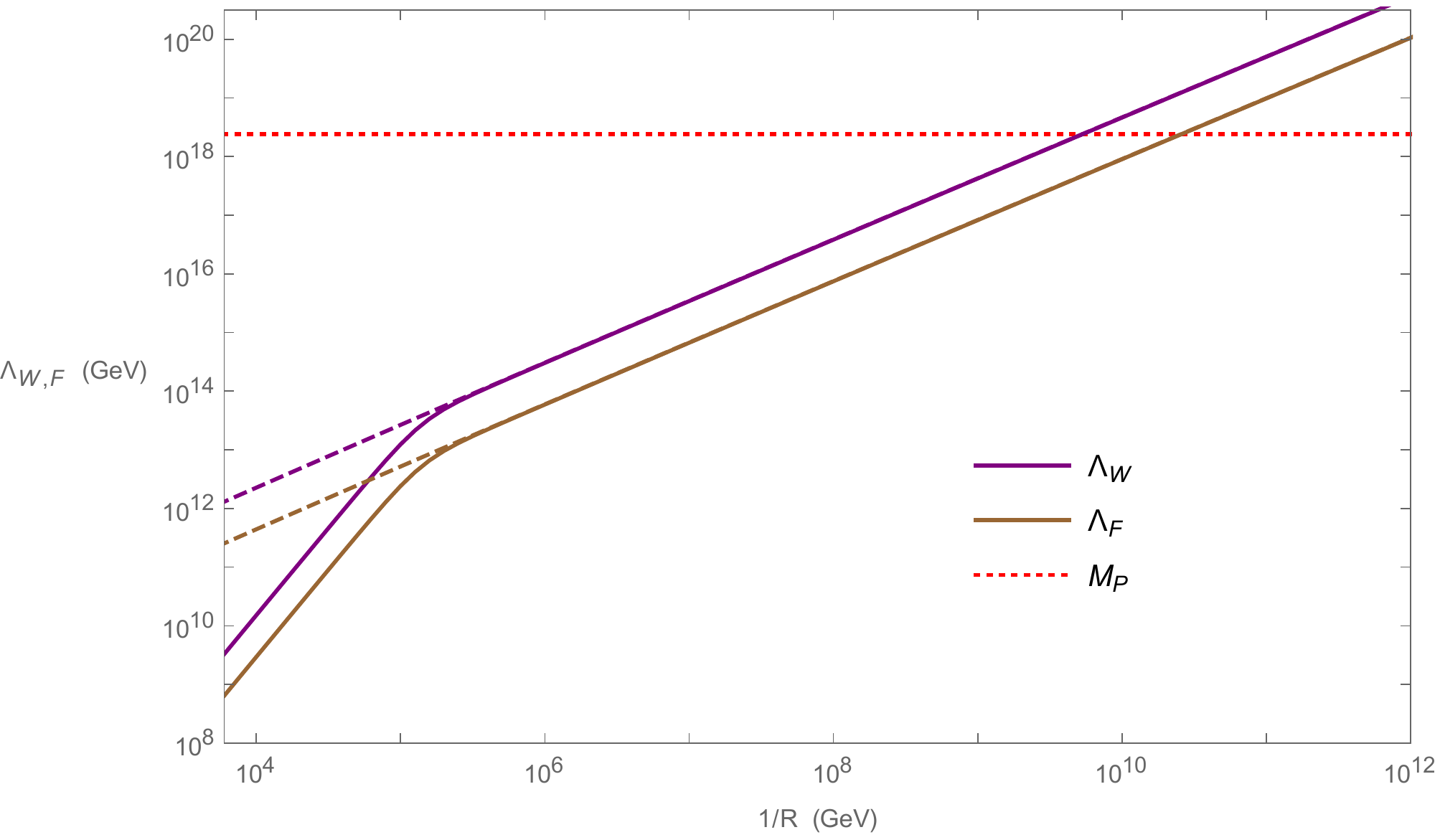}
\caption{Lower limit on the effective scale 
of the dimension six Weinberg (four-fermion) operator, depicted in purple (orange) as a function of the extra dimension scale $1/R$, assuming $\epsilon=0.30$. The Planck scale is shown as a dotted line for reference. 
The dashed lines represent the limit from the approximation \eqref{eq:approx5Dtheta}. The deviation from \eqref{eq:approx5Dtheta} arises since for small $1/R$, large QCD instantons begin to dominate the instanton integral $\chi(0)$.}  
\label{fig:6limit}
\end{figure}

Using an approximate expression for the integrals in \eqref{eq:topsuscint} and \eqref{eq:Wmixed} with the effective action \eqref{eff_action}, the induced $\theta$ from 5D small instantons is
\begin{equation}
    \theta_{\rm ind}\approx
    \xi_{W,F}\frac{\Lambda_{\rm SI}^2}{\Lambda_{W,F}^2}~,
    \label{eq:approx5Dtheta}
\end{equation}
where $\xi_{W}$ and $\xi_F$ are defined under \eqref{eq:QCDindth} and \eqref{eq:SIindth_product}, respectively.
The induced $\theta$ no longer necessarily decouples in the limit $\Lambda_{\rm SI},\Lambda_{W,F}\rightarrow \infty$. Imposing the constraint \eqref{eq:indthlimit} leads to the limit $\Lambda_{\rm SI}/\Lambda_{W}(\Lambda_{F}) \lesssim 
10^{-7}(10^{-6})$. For $\Lambda_{W}(\Lambda_{F})  = M_P$ this implies an upper bound $\Lambda_{\rm SI} \lesssim 10^{11}(10^{12})$~GeV on the 5D strong coupling scale.
The limit on $\Lambda_{W,F}$ from an exact numerical evaluation of $\theta_{\rm ind}$ is shown in Figure~\ref{fig:6limit}. We see that the limit on $\Lambda_{W,F}$ deviates from \eqref{eq:approx5Dtheta} for small $1/R$ (and hence small $\Lambda_5$).
The limits on the ratio $\Lambda_{\rm SI}/\Lambda_{W,F}$ imply that for the case when $\Lambda_{W,F}\sim \Lambda_5(=\Lambda_{\rm SI})$, the dimension six terms would need to be generated {from some new physics} in the UV completion of the 5D model with an additional suppression in the otherwise order-one coefficients.

\begin{figure}[h]
    \centering
     \includegraphics[width=0.8\textwidth]{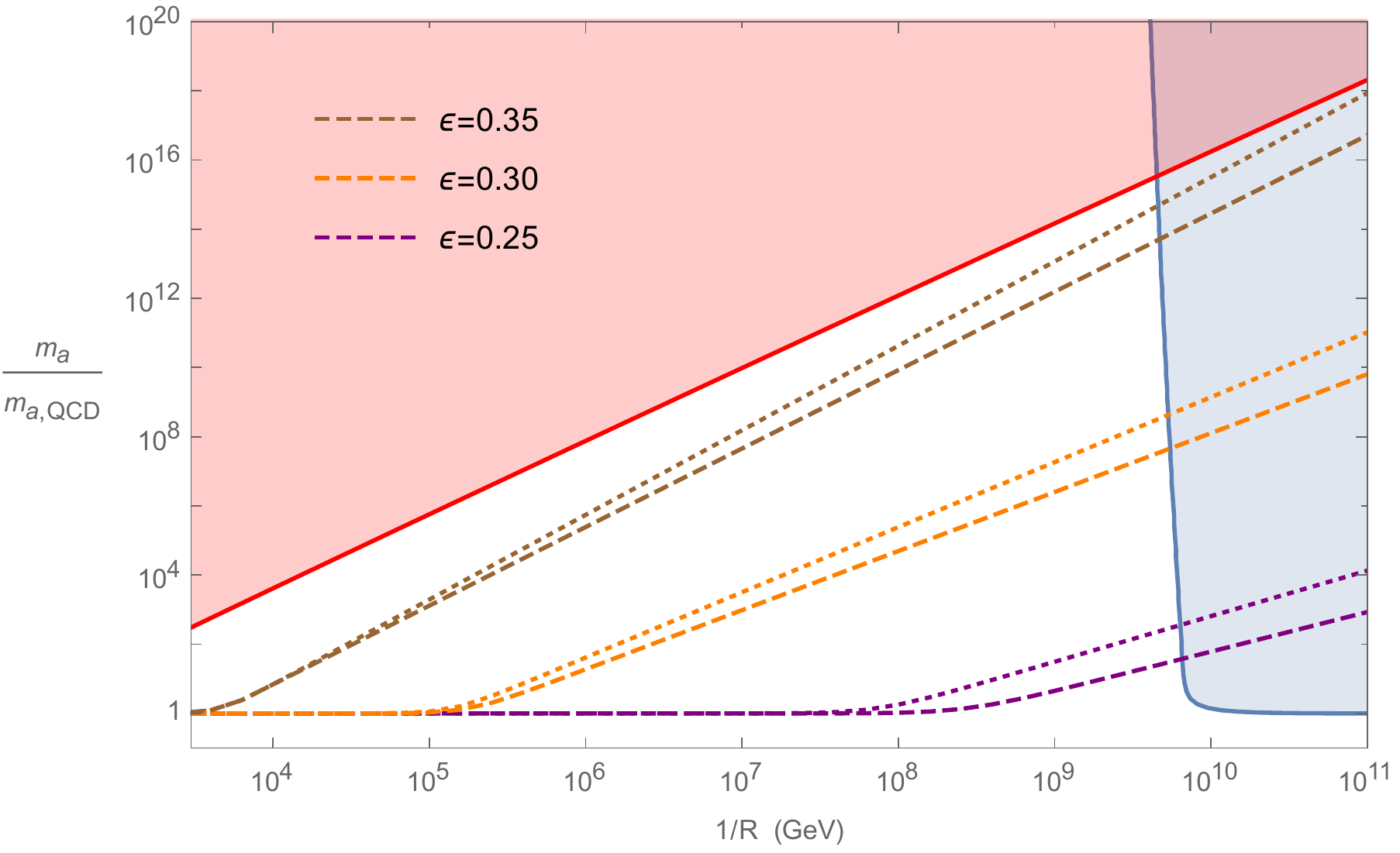}
    \caption{The ratio of the enhanced axion mass to  the QCD axion mass as a function of the extra dimension scale, $1/R$. The dotted contour lines assume $f_a>\Lambda_{\rm SI}$ and depict the ratio for different values of the perturbitivity parameter $\epsilon$, up to the maximum possible enhancement in the red shaded region. The dashed contours assume $f_a=10^6$~GeV and include the suppression \eqref{eq:topsussuppress} when $f_a< \Lambda_{\rm SI}$.
    The blue shaded region to the right shows the excluded $1/R$ range due to the Weinberg gluonic operator.}
    \label{fig:axion5Dmass}
\end{figure}

The corresponding range of axion mass enhancement is depicted in Figure~\ref{fig:axion5Dmass}.
Note that both effects of small instantons -- the enhancement of the axion mass and the shift in the axion potential minimum due to $CP$-violating operators -- are dominant only for large $1/R$, since eventually large (QCD) instantons dominate the susceptibility at small values of $1/R$.

{ 
Furthermore, when $f_a< \Lambda_{\rm SI}$ the axion mass enhancement is reduced by the factor $(f_a/\Lambda_{\rm SI})^{1/6}$ as obtained from \eqref{eq:topsussuppress}. This means that in the experimentally viable region of  
 $ m_a \gtrsim 100$~MeV and $f_a\lesssim 10^7$~GeV~\cite{Cadamuro:2011fd, Jaeckel:2015jla}, the axion mass enhancement is reduced by up to an order of magnitude, as can be seen in Figure \ref{fig:axion5Dmass}, where we have taken $f_a=10^6$~GeV as a representative value. 
}

\subsection{Enhanced EDMs}
Compared to QCD, the small instanton contributions provide an enhancement to the EDMs due to $CP$-violating sources. In particular, using \eqref{eq:SIindth_product}, \eqref{eq:QCDFindth} and \eqref{eq:approx5Dtheta} we see that the neutron EDM \eqref{dn} is enhanced by a factor of $\Lambda_{\rm SI}^2/\Lambda_{\rm QCD}^2$ compared to the $\theta$ induced from new $CP$-odd sources in QCD (see \eqref{eq:QCDindth}). Therefore, measuring the neutron EDM can be interpreted as a probe of the small instanton scale, $\Lambda_{\rm SI}$. 
For example, if $\Lambda_{W,F} = M_P$, this corresponds to modified strong dynamics at scales of order $\Lambda_{\rm SI}\sim 10^{8}-10^{11}$~GeV, where the lower limit represents a neutron EDM value equivalent to the Standard Model CKM contribution. Furthermore, if the $CP$-violating sources appear at scales lower than the Planck scale, then any new contribution due to small instantons will appear at even lower scales $10^4\text{~GeV}\lesssim\Lambda_{\rm SI}\lesssim 10^8$~GeV, where the model-dependent lower limit corresponds to the scale of axion mass enhancement.

Finally note that when $f_a\lesssim \Lambda_{\rm SI}$, the UV completion of the dimension five axion-gluon coupling does not affect the predictions for the induced $\theta$. Since the neutron EDM \eqref{dn} depends only on the ratio of the mixed correlators with the topological susceptibility, the suppression factor in \eqref{eq:topsussuppress} cancels, leaving the results for the induced $\theta$ unchanged.

\section{Conclusion}\label{sec4}

The QCD axion solution provides an elegant mechanism for solving the strong $CP$ problem in such a way that an arbitrarily large amount of $CP$ violation at UV scales $\Lambda_{\rm CP}$ can be sufficiently decoupled as $\Lambda_{\rm CP}\rightarrow \infty$. This is in contrast 
with solutions to the strong $CP$ problem that invoke exact discrete symmetries. For these solutions there is a nondecoupling of the additional sources of $CP$ violation, which means that arbitrarily large amounts of $CP$ violation cannot be tolerated at UV scales in models with exact parity or $CP$ symmetry.

Heavy axion models represent a qualitatively different class of solution to the strong $CP$ problem in which new dynamics at some UV scale $\Lambda_{\rm SI}$ magnifies the effect of small instantons (which are normally exponentially suppressed), giving rise to a new contribution and enhancement of the axion mass. This has led to renewed interest in axion searches outside the usual QCD axion mass window. However, in the presence of additional sources of $CP$- violation, the enhanced effect of small instantons could also lead to enhanced EDM observables such as the neutron EDM as well as possible nondecoupling effects.

We have estimated these effects by calculating the topological susceptibility and mixed correlators in the presence of two $CP$-violating dimension six operators: the Weinberg gluonic operator and a $CP$-odd four-fermion operator. The calculation is performed using an instanton (or anti-instanton) background where Standard Model fermion chiral zero modes in the 't Hooft vertex are closed with the Higgs boson.
Identifying the scale of the additional sources of $CP$ violation with $\Lambda_{\rm CP}$ we find that the axion potential minimum shifts by an amount $\theta_{\rm ind}\propto \Lambda^2_{\rm SI}/\Lambda^2_{\rm CP}$ in several heavy axion models, where $\Lambda_{\rm SI}$ is the scale where small instanton effects dominate. This result reveals that unlike the minimal QCD axion models, the amount of decoupling is limited, although not as restrictive as models with exact discrete symmetries.
Imposing the neutron EDM derived limit $|{\bar\theta}|\lesssim 10^{-10}$, we obtain the constraint $\Lambda_{\rm SI}/\Lambda_{\rm CP}\lesssim 10^{-8}$, which is stronger than the naive estimate of $10^{-5}$ due to sizable prefactors that depend on the particular heavy axion model.
In particular, for a benchmark value of $\Lambda_{\rm CP}\simeq M_P$ requires $\Lambda_{\rm SI}\lesssim 10^{10}$ GeV (as can be seen in Figure~\ref{fig:axion5Dmass} for the 5D small instanton model).

The modification of the decoupling behavior is a direct consequence of the new dynamical scale $\Lambda_{\rm SI}$. Our results therefore imply that EDM observables such as the neutron EDM can be enhanced in heavy axion models up to the current experimental limit $d_n\lesssim 10^{-26}$~e$\cdot$cm. This compares with the SM CKM prediction ($\sim 10^{-32}-10^{-31}$~e$\cdot$cm). Thus, besides axion searches, EDM observables provide another probe of UV scales in heavy axion models associated with new dynamics, assuming that this class of models plays any role in solving the strong $CP$ problem.

\section*{Acknowledgments}
T.G. thanks Raymond Co and Keisuke Harigaya for useful discussions.
This work is supported in part by the Department of Energy under Grant No.~DE-SC0011842 at the University of Minnesota. T.G. is also supported by the Simons Foundation. 
\bibliographystyle{JHEP}
\bibliography{references}

\providecommand{\href}[2]{#2}\begingroup\raggedright\begin{thebibliography}{10}

\bibitem{Kobayashi:1973fv}
M.~Kobayashi and T.~Maskawa, \emph{{CP Violation in the Renormalizable Theory
  of Weak Interaction}}, \href{https://doi.org/10.1143/PTP.49.652}{\emph{Prog.
  Theor. Phys.} {\bfseries 49} (1973) 652}.

\bibitem{Abel:2020pzs}
C.~Abel et~al., \emph{{Measurement of the Permanent Electric Dipole Moment of
  the Neutron}},
  \href{https://doi.org/10.1103/PhysRevLett.124.081803}{\emph{Phys. Rev. Lett.}
  {\bfseries 124} (2020) 081803}
  [\href{https://arxiv.org/abs/2001.11966}{{\ttfamily 2001.11966}}].

\bibitem{Graner:2016ses}
B.~Graner, Y.~Chen, E.G.~Lindahl and B.R.~Heckel, \emph{{Reduced Limit on the
  Permanent Electric Dipole Moment of Hg199}},
  \href{https://doi.org/10.1103/PhysRevLett.116.161601}{\emph{Phys. Rev. Lett.}
  {\bfseries 116} (2016) 161601}
  [\href{https://arxiv.org/abs/1601.04339}{{\ttfamily 1601.04339}}].

\bibitem{Peccei:1977hh}
R.D.~Peccei and H.R.~Quinn, \emph{{CP Conservation in the Presence of
  Instantons}}, \href{https://doi.org/10.1103/PhysRevLett.38.1440}{\emph{Phys.
  Rev. Lett.} {\bfseries 38} (1977) 1440}.

\bibitem{Weinberg:1977ma}
S.~Weinberg, \emph{{A New Light Boson?}},
  \href{https://doi.org/10.1103/PhysRevLett.40.223}{\emph{Phys. Rev. Lett.}
  {\bfseries 40} (1978) 223}.

\bibitem{Wilczek:1977pj}
F.~Wilczek, \emph{{Problem of Strong $P$ and $T$ Invariance in the Presence of
  Instantons}}, \href{https://doi.org/10.1103/PhysRevLett.40.279}{\emph{Phys.
  Rev. Lett.} {\bfseries 40} (1978) 279}.

\bibitem{Kim:1979if}
J.E.~Kim, \emph{{Weak Interaction Singlet and Strong CP Invariance}},
  \href{https://doi.org/10.1103/PhysRevLett.43.103}{\emph{Phys. Rev. Lett.}
  {\bfseries 43} (1979) 103}.

\bibitem{Shifman:1979if}
M.A.~Shifman, A.I.~Vainshtein and V.I.~Zakharov, \emph{{Can Confinement Ensure
  Natural CP Invariance of Strong Interactions?}},
  \href{https://doi.org/10.1016/0550-3213(80)90209-6}{\emph{Nucl. Phys. B}
  {\bfseries 166} (1980) 493}.

\bibitem{Dine:1981rt}
M.~Dine, W.~Fischler and M.~Srednicki, \emph{{A Simple Solution to the Strong
  CP Problem with a Harmless Axion}},
  \href{https://doi.org/10.1016/0370-2693(81)90590-6}{\emph{Phys. Lett.}
  {\bfseries 104B} (1981) 199}.

\bibitem{Zhitnitsky:1980tq}
A.R.~Zhitnitsky, \emph{{On Possible Suppression of the Axion Hadron
  Interactions. (In Russian)}}, {\emph{Sov. J. Nucl. Phys.} {\bfseries 31}
  (1980) 260}.

\bibitem{Bigi:1990kz}
I.I.Y.~Bigi and N.G.~Uraltsev, \emph{{Induced Multi - Gluon Couplings and the
  Neutron Electric Dipole Moment}},
  \href{https://doi.org/10.1016/0550-3213(91)90339-Y}{\emph{Nucl. Phys. B}
  {\bfseries 353} (1991) 321}.

\bibitem{Pospelov:2000bw}
M.~Pospelov and A.~Ritz, \emph{{Neutron EDM from electric and chromoelectric
  dipole moments of quarks}},
  \href{https://doi.org/10.1103/PhysRevD.63.073015}{\emph{Phys. Rev. D}
  {\bfseries 63} (2001) 073015}
  [\href{https://arxiv.org/abs/hep-ph/0010037}{{\ttfamily hep-ph/0010037}}].

\bibitem{Pospelov:2005pr}
M.~Pospelov and A.~Ritz, \emph{{Electric dipole moments as probes of new
  physics}}, \href{https://doi.org/10.1016/j.aop.2005.04.002}{\emph{Annals
  Phys.} {\bfseries 318} (2005) 119}
  [\href{https://arxiv.org/abs/hep-ph/0504231}{{\ttfamily hep-ph/0504231}}].

\bibitem{Nelson:1983zb}
A.E.~Nelson, \emph{{Naturally Weak CP Violation}},
  \href{https://doi.org/10.1016/0370-2693(84)92025-2}{\emph{Phys. Lett. B}
  {\bfseries 136} (1984) 387}.

\bibitem{Barr:1984qx}
S.M.~Barr, \emph{{Solving the Strong CP Problem Without the Peccei-Quinn
  Symmetry}}, \href{https://doi.org/10.1103/PhysRevLett.53.329}{\emph{Phys.
  Rev. Lett.} {\bfseries 53} (1984) 329}.

\bibitem{Babu:1988mw}
K.S.~Babu and R.N.~Mohapatra, \emph{{{CP} Violation in Seesaw Models of Quark
  Masses}}, \href{https://doi.org/10.1103/PhysRevLett.62.1079}{\emph{Phys. Rev.
  Lett.} {\bfseries 62} (1989) 1079}.

\bibitem{Babu:1989rb}
K.S.~Babu and R.N.~Mohapatra, \emph{{A Solution to the Strong {CP} Problem
  Without an Axion}},
  \href{https://doi.org/10.1103/PhysRevD.41.1286}{\emph{Phys. Rev. D}
  {\bfseries 41} (1990) 1286}.

\bibitem{Mohapatra:1995xd}
R.N.~Mohapatra and A.~Rasin, \emph{{Simple supersymmetric solution to the
  strong CP problem}},
  \href{https://doi.org/10.1103/PhysRevLett.76.3490}{\emph{Phys. Rev. Lett.}
  {\bfseries 76} (1996) 3490}
  [\href{https://arxiv.org/abs/hep-ph/9511391}{{\ttfamily hep-ph/9511391}}].

\bibitem{Kuchimanchi:1995rp}
R.~Kuchimanchi, \emph{{Solution to the strong CP problem: Supersymmetry with
  parity}}, \href{https://doi.org/10.1103/PhysRevLett.76.3486}{\emph{Phys. Rev.
  Lett.} {\bfseries 76} (1996) 3486}
  [\href{https://arxiv.org/abs/hep-ph/9511376}{{\ttfamily hep-ph/9511376}}].

\bibitem{Holdom:1999ny}
B.~Holdom, \emph{{Nonstandard order parameters and the origin of CP
  violation}}, \href{https://doi.org/10.1103/PhysRevD.61.011702}{\emph{Phys.
  Rev. D} {\bfseries 61} (1999) 011702}
  [\href{https://arxiv.org/abs/hep-ph/9907361}{{\ttfamily hep-ph/9907361}}].

\bibitem{Hiller:2001qg}
G.~Hiller and M.~Schmaltz, \emph{{Solving the Strong CP Problem with
  Supersymmetry}},
  \href{https://doi.org/10.1016/S0370-2693(01)00814-0}{\emph{Phys. Lett. B}
  {\bfseries 514} (2001) 263}
  [\href{https://arxiv.org/abs/hep-ph/0105254}{{\ttfamily hep-ph/0105254}}].

\bibitem{Berezhiani:2000gh}
Z.~Berezhiani, L.~Gianfagna and M.~Giannotti, \emph{{Strong CP problem and
  mirror world: The Weinberg-Wilczek axion revisited}},
  \href{https://doi.org/10.1016/S0370-2693(00)01392-7}{\emph{Phys. Lett.}
  {\bfseries B500} (2001) 286}
  [\href{https://arxiv.org/abs/hep-ph/0009290}{{\ttfamily hep-ph/0009290}}].

\bibitem{Hook:2014cda}
A.~Hook, \emph{{Anomalous solutions to the strong CP problem}},
  \href{https://doi.org/10.1103/PhysRevLett.114.141801}{\emph{Phys. Rev. Lett.}
  {\bfseries 114} (2015) 141801}
  [\href{https://arxiv.org/abs/1411.3325}{{\ttfamily 1411.3325}}].

\bibitem{Ellis:1978hq}
J.R.~Ellis and M.K.~Gaillard, \emph{{Strong and Weak CP Violation}},
  \href{https://doi.org/10.1016/0550-3213(79)90297-9}{\emph{Nucl. Phys. B}
  {\bfseries 150} (1979) 141}.

\bibitem{Khriplovich:1985jr}
I.B.~Khriplovich, \emph{{Quark Electric Dipole Moment and Induced $\theta$ Term
  in the {Kobayashi-Maskawa} Model}},
  \href{https://doi.org/10.1016/0370-2693(86)90245-5}{\emph{Phys. Lett. B}
  {\bfseries 173} (1986) 193}.

\bibitem{Pospelov:1996be}
M.E.~Pospelov, \emph{{Radiative corrections to theta term in the left-right
  supersymmetric models}},
  \href{https://doi.org/10.1016/S0370-2693(96)01467-0}{\emph{Phys. Lett. B}
  {\bfseries 391} (1997) 324}
  [\href{https://arxiv.org/abs/hep-ph/9609458}{{\ttfamily hep-ph/9609458}}].

\bibitem{Frampton:1996vxg}
P.H.~Frampton and O.C.W.~Kong, \emph{{Strong CP and low-energy supersymmetry}},
  \href{https://doi.org/10.1016/S0370-2693(97)00245-1}{\emph{Phys. Lett. B}
  {\bfseries 402} (1997) 297}
  [\href{https://arxiv.org/abs/hep-ph/9612452}{{\ttfamily hep-ph/9612452}}].

\bibitem{deVries:2021pzl}
J.~de~Vries, P.~Draper and H.H.~Patel, \emph{{Do Minimal Parity Solutions to
  the Strong $CP$ Problem Work?}},
  \href{https://arxiv.org/abs/2109.01630}{{\ttfamily 2109.01630}}.

\bibitem{Dimopoulos:1979pp}
S.~Dimopoulos, \emph{{A Solution of the Strong {CP} Problem in Models With
  Scalars}}, \href{https://doi.org/10.1016/0370-2693(79)91233-4}{\emph{Phys.
  Lett. B} {\bfseries 84} (1979) 435}.

\bibitem{Rubakov:1997vp}
V.A.~Rubakov, \emph{{Grand unification and heavy axion}},
  \href{https://doi.org/10.1134/1.567390}{\emph{JETP Lett.} {\bfseries 65}
  (1997) 621} [\href{https://arxiv.org/abs/hep-ph/9703409}{{\ttfamily
  hep-ph/9703409}}].

\bibitem{Fukuda:2015ana}
H.~Fukuda, K.~Harigaya, M.~Ibe and T.T.~Yanagida, \emph{{Model of visible QCD
  axion}}, \href{https://doi.org/10.1103/PhysRevD.92.015021}{\emph{Phys. Rev.}
  {\bfseries D92} (2015) 015021}
  [\href{https://arxiv.org/abs/1504.06084}{{\ttfamily 1504.06084}}].

\bibitem{Gherghetta:2016fhp}
T.~Gherghetta, N.~Nagata and M.~Shifman, \emph{{A Visible QCD Axion from an
  Enlarged Color Group}},
  \href{https://doi.org/10.1103/PhysRevD.93.115010}{\emph{Phys. Rev.}
  {\bfseries D93} (2016) 115010}
  [\href{https://arxiv.org/abs/1604.01127}{{\ttfamily 1604.01127}}].

\bibitem{Gherghetta:2020ofz}
T.~Gherghetta and M.D.~Nguyen, \emph{{A Composite Higgs with a Heavy Composite
  Axion}}, \href{https://doi.org/10.1007/JHEP12(2020)094}{\emph{JHEP}
  {\bfseries 12} (2020) 094}
  [\href{https://arxiv.org/abs/2007.10875}{{\ttfamily 2007.10875}}].

\bibitem{Hook:2019qoh}
A.~Hook, S.~Kumar, Z.~Liu and R.~Sundrum, \emph{{High Quality QCD Axion and the
  LHC}}, \href{https://doi.org/10.1103/PhysRevLett.124.221801}{\emph{Phys. Rev.
  Lett.} {\bfseries 124} (2020) 221801}
  [\href{https://arxiv.org/abs/1911.12364}{{\ttfamily 1911.12364}}].

\bibitem{Holdom:1982ex}
B.~Holdom and M.E.~Peskin, \emph{{Raising the Axion Mass}},
  \href{https://doi.org/10.1016/0550-3213(82)90228-0}{\emph{Nucl. Phys.}
  {\bfseries B208} (1982) 397}.

\bibitem{Holdom:1985vx}
B.~Holdom, \emph{{Strong QCD at High-energies and a Heavy Axion}},
  \href{https://doi.org/10.1016/0370-2693(85)90371-5}{\emph{Phys. Lett.}
  {\bfseries 154B} (1985) 316}.

\bibitem{Dine:1986bg}
M.~Dine and N.~Seiberg, \emph{{String Theory and the Strong {CP} Problem}},
  \href{https://doi.org/10.1016/0550-3213(86)90043-X}{\emph{Nucl. Phys.}
  {\bfseries B273} (1986) 109}.

\bibitem{Flynn:1987rs}
J.M.~Flynn and L.~Randall, \emph{{A Computation of the Small Instanton
  Contribution to the Axion Potential}},
  \href{https://doi.org/10.1016/0550-3213(87)90089-7}{\emph{Nucl. Phys.}
  {\bfseries B293} (1987) 731}.

\bibitem{Agrawal:2017ksf}
P.~Agrawal and K.~Howe, \emph{{Factoring the Strong CP Problem}},
  \href{https://doi.org/10.1007/JHEP12(2018)029}{\emph{JHEP} {\bfseries 12}
  (2018) 029} [\href{https://arxiv.org/abs/1710.04213}{{\ttfamily
  1710.04213}}].

\bibitem{Csaki:2019vte}
C.~Cs\'aki, M.~Ruhdorfer and Y.~Shirman, \emph{{UV Sensitivity of the Axion
  Mass from Instantons in Partially Broken Gauge Groups}},
  \href{https://doi.org/10.1007/JHEP04(2020)031}{\emph{JHEP} {\bfseries 04}
  (2020) 031} [\href{https://arxiv.org/abs/1912.02197}{{\ttfamily
  1912.02197}}].

\bibitem{Gherghetta:2020keg}
T.~Gherghetta, V.V.~Khoze, A.~Pomarol and Y.~Shirman, \emph{{The Axion Mass
  from 5D Small Instantons}},
  \href{https://doi.org/10.1007/JHEP03(2020)063}{\emph{JHEP} {\bfseries 03}
  (2020) 063} [\href{https://arxiv.org/abs/2001.05610}{{\ttfamily
  2001.05610}}].

\bibitem{Gherghetta:2021jnn}
T.~Gherghetta and A.~Pomarol, \emph{{Small instantons in weakly-gauged
  holographic models}},
  \href{https://doi.org/10.1007/JHEP11(2021)136}{\emph{JHEP} {\bfseries 11}
  (2021) 136} [\href{https://arxiv.org/abs/2110.01762}{{\ttfamily
  2110.01762}}].

\bibitem{Kitano:2021fdl}
R.~Kitano and W.~Yin, \emph{{Strong CP problem and axion dark matter with small
  instantons}}, \href{https://doi.org/10.1007/JHEP07(2021)078}{\emph{JHEP}
  {\bfseries 07} (2021) 078}
  [\href{https://arxiv.org/abs/2103.08598}{{\ttfamily 2103.08598}}].

\bibitem{Belavin:1975fg}
A.A.~Belavin, A.M.~Polyakov, A.S.~Schwartz and Y.S.~Tyupkin,
  \emph{{Pseudoparticle Solutions of the Yang-Mills Equations}},
  \href{https://doi.org/10.1016/0370-2693(75)90163-X}{\emph{Phys. Lett. B}
  {\bfseries 59} (1975) 85}.

\bibitem{tHooft:1976snw}
G.~'t~Hooft, \emph{{Computation of the Quantum Effects Due to a
  Four-Dimensional Pseudoparticle}},
  \href{https://doi.org/10.1103/PhysRevD.18.2199.3,
  10.1103/PhysRevD.14.3432}{\emph{Phys. Rev.} {\bfseries D14} (1976) 3432}.

\bibitem{Jackiw1977ConformalConfigurations}
R.~Jackiw, C.~Nohl and C.~Rebbi, \emph{{Conformal properties of pseudoparticle
  configurations}},
  \href{https://doi.org/10.1103/PhysRevD.15.1642}{\emph{Physical Review D}
  {\bfseries 15} (1977) 1642}.

\bibitem{Witten:1979vv}
E.~Witten, \emph{{Current Algebra Theorems for the U(1) Goldstone Boson}},
  \href{https://doi.org/10.1016/0550-3213(79)90031-2}{\emph{Nucl. Phys. B}
  {\bfseries 156} (1979) 269}.

\bibitem{Cordes:1985um}
S.F.~Cordes, \emph{{The Instanton Induced Superpotential in Supersymmetric
  {QCD}}}, \href{https://doi.org/10.1016/0550-3213(86)90381-0}{\emph{Nucl.
  Phys. B} {\bfseries 273} (1986) 629}.

\bibitem{Vainshtein:1981wh}
A.I.~Vainshtein, V.I.~Zakharov, V.A.~Novikov and M.A.~Shifman, \emph{{ABC's of
  Instantons}},
  \href{https://doi.org/10.1070/PU1982v025n04ABEH004533}{\emph{Sov. Phys. Usp.}
  {\bfseries 25} (1982) 195}.

\bibitem{Callan:1977gz}
C.G.~Callan, Jr., R.F.~Dashen and D.J.~Gross, \emph{{Toward a Theory of the
  Strong Interactions}},
  \href{https://doi.org/10.1103/PhysRevD.17.2717}{\emph{Phys. Rev.} {\bfseries
  D17} (1978) 2717}.

\bibitem{Diakonov:1983hh}
D.~Diakonov and V.Y.~Petrov, \emph{{Instanton Based Vacuum from Feynman
  Variational Principle}},
  \href{https://doi.org/10.1016/0550-3213(84)90432-2}{\emph{Nucl. Phys. B}
  {\bfseries 245} (1984) 259}.

\bibitem{Shuryak:1988rf}
E.V.~Shuryak, \emph{{Instantons in {QCD}. 1. Properties of the 'Instanton
  Liquid'}}, \href{https://doi.org/10.1016/0550-3213(89)90618-4}{\emph{Nucl.
  Phys. B} {\bfseries 319} (1989) 521}.

\bibitem{Diakonov:1985eg}
D.~Diakonov and V.Y.~Petrov, \emph{{A Theory of Light Quarks in the Instanton
  Vacuum}}, \href{https://doi.org/10.1016/0550-3213(86)90011-8}{\emph{Nucl.
  Phys. B} {\bfseries 272} (1986) 457}.

\bibitem{Diakonov:1995qy}
D.~Diakonov, M.V.~Polyakov and C.~Weiss, \emph{{Hadronic matrix elements of
  gluon operators in the instanton vacuum}},
  \href{https://doi.org/10.1016/0550-3213(95)00675-3}{\emph{Nucl. Phys. B}
  {\bfseries 461} (1996) 539}
  [\href{https://arxiv.org/abs/hep-ph/9510232}{{\ttfamily hep-ph/9510232}}].

\bibitem{Diakonov:1984vw}
D.~Diakonov and V.Y.~Petrov, \emph{{Chiral Condensate in the Instanton
  Vacuum}}, \href{https://doi.org/10.1016/0370-2693(84)90132-1}{\emph{Phys.
  Lett. B} {\bfseries 147} (1984) 351}.

\bibitem{Weinberg:1989dx}
S.~Weinberg, \emph{{Larger Higgs Exchange Terms in the Neutron Electric Dipole
  Moment}}, \href{https://doi.org/10.1103/PhysRevLett.63.2333}{\emph{Phys. Rev.
  Lett.} {\bfseries 63} (1989) 2333}.

\bibitem{Weiss:2021kpt}
C.~Weiss, \emph{{Nucleon matrix element of Weinberg's CP-odd gluon operator
  from the instanton vacuum}},
  \href{https://doi.org/10.1016/j.physletb.2021.136447}{\emph{Phys. Lett. B}
  {\bfseries 819} (2021) 136447}
  [\href{https://arxiv.org/abs/2103.13471}{{\ttfamily 2103.13471}}].

\bibitem{Bonnefoy:2020llz}
Q.~Bonnefoy, P.~Cox, E.~Dudas, T.~Gherghetta and M.D.~Nguyen, \emph{{Flavoured
  Warped Axion}}, \href{https://doi.org/10.1007/JHEP04(2021)084}{\emph{JHEP}
  {\bfseries 04} (2021) 084}
  [\href{https://arxiv.org/abs/2012.09728}{{\ttfamily 2012.09728}}].

\bibitem{Pospelov:2005ks}
M.~Pospelov, A.~Ritz and Y.~Santoso, \emph{{Flavor and CP violating physics
  from new supersymmetric thresholds}},
  \href{https://doi.org/10.1103/PhysRevLett.96.091801}{\emph{Phys. Rev. Lett.}
  {\bfseries 96} (2006) 091801}
  [\href{https://arxiv.org/abs/hep-ph/0510254}{{\ttfamily hep-ph/0510254}}].

\bibitem{An:2009zh}
H.~An, X.~Ji and F.~Xu, \emph{{P-odd and CP-odd Four-Quark Contributions to
  Neutron EDM}}, \href{https://doi.org/10.1007/JHEP02(2010)043}{\emph{JHEP}
  {\bfseries 02} (2010) 043} [\href{https://arxiv.org/abs/0908.2420}{{\ttfamily
  0908.2420}}].

\bibitem{Cadamuro:2011fd}
D.~Cadamuro and J.~Redondo, \emph{{Cosmological bounds on pseudo
  Nambu-Goldstone bosons}},
  \href{https://doi.org/10.1088/1475-7516/2012/02/032}{\emph{JCAP} {\bfseries
  02} (2012) 032} [\href{https://arxiv.org/abs/1110.2895}{{\ttfamily
  1110.2895}}].

\bibitem{Jaeckel:2015jla}
J.~Jaeckel and M.~Spannowsky, \emph{{Probing MeV to 90 GeV axion-like particles
  with LEP and LHC}},
  \href{https://doi.org/10.1016/j.physletb.2015.12.037}{\emph{Phys. Lett. B}
  {\bfseries 753} (2016) 482}
  [\href{https://arxiv.org/abs/1509.00476}{{\ttfamily 1509.00476}}].

\bibitem{Co:2022aav}
R.T.~Co, T.~Gherghetta and K.~Harigaya, \emph{{Axiogenesis with a Heavy QCD
  Axion}},  \href{https://arxiv.org/abs/2206.00678}{{\ttfamily 2206.00678}}.

\end{thebibliography}\endgroup

\end{document}